# Ergodicity-breaking reveals time optimal decision making in humans

David Meder[1], Finn Rabe[1,2], Tobias Morville[1], Kristoffer H. Madsen[1,3], Magnus T. Koudahl[1,4], Ray J. Dolan[6] Hartwig R. Siebner[1,7,8], Oliver J. Hulme[1]*

[1]Danish Research Centre for Magnetic Resonance, Centre for Functional and Diagnostic Imaging and Research, Copenhagen University Hospital Hvidovre, Kettegard Allé 30, 2650, Hvidovre, Denmark. [2]Neural Control of Movement Lab, Winterthurerstrasse 190 Building Y36, ETH Zurich, 8057, Zurich, Switzerland. [3] Department of Applied Mathematics and Computer Science, Technical University of Denmark, 2800, Kongens Lyngby, Denmark. [4]Department of Electrical Engineering, Eindhoven University of Technology, 6500 MB, Eindhoven, the Netherlands. [6] Max Planck UCL Centre for Computational Psychiatry and Ageing Research, 10-12 Russell Square, London, WC1B 5EH, United Kingdom. [7]Department of Neurology, Copenhagen University Hospital Bispebjerg, Bispebjerg Bakke 23, 2400 Copenhagen, Denmark. [8]Institute for Clinical Medicine, Faculty of Medical and Health Sciences, University of Copenhagen, Blegdamsvej 9, 2100 Copenhagen, Denmark. *corresponding author: oliverh@drcmr.dk

ORCID ID: D. Meder: 0000-0001-9689-0869, T. Morville: 0000-0003-1079-9891, K. H. Madsen: 0000- 0001-8606-7641, R. J. Dolan: 0000-0001-9356-761X, H. R. Siebner: 0000-0002-3756-9431, O. J. Hulme: 0000-0003-3139-4324

**Ergodicity describes an equivalence between the expectation value and the time average of observables. Applied to human behaviour, ergodic theories of decision-making reveal how individuals should tolerate risk in different environments. To optimise wealth over time, agents should adapt their utility function according to the dynamical setting they face. Linear utility is optimal for additive dynamics, whereas logarithmic utility is optimal for multiplicative dynamics. Whether humans approximate time optimal behavior across different dynamics is unknown. Here we compare the effects of additive versus multiplicative gamble dynamics on risky choice. We show that utility functions are modulated by gamble dynamics in ways not explained by prevailing decision theory. Instead, as predicted by time optimality, risk aversion increases under multiplicative dynamics, distributing close to the values that maximise the time average growth of wealth. We suggest that our findings motivate a need for explicitly grounding theories of decision-making on ergodic considerations.**

Ergodicity is a foundational concept in models of physical systems that include elements of randomness[1,2,3]. A physical observable is ergodic if the average over its possible states, is the same as its average over time. For instance, the velocity of gas molecules in a chamber is ergodic if averaging over all molecules at a fixed time (an expectation value) yields the same value, as averaging a single molecule over an extended period of time (a time average). In other words, ergodicity ensures an equality between the time average and the expectation value. The relevance of ergodicity to human behavior is that it provides important constraints for thinking about how agents should compute averages when making decisions[4,5].

In the behavioral sciences, decision making is studied predominantly using experiments with additive dynamics, where choice outcomes exert additive effects on wealth. An agent might gamble on a coin toss for a gain of $1 each time they win, they might score a point each time they correctly execute a motor action, and so on. In

these examples, changes in wealth are ergodic, and in such settings a linear utility function is optimal for maximising the growth of wealth over time[5]. In other words, for this utility function, when changes in expected utility are maximized per unit time, this maximizes the time average growth rate of wealth (Fig. 1f). However, not all dynamics individuals face are additive. Some dynamics in the environment are multiplicative, for instance. Examples of multiplicative dynamics include stock market investments, or the compound interests on savings, and the spread of infectious diseases. Settings with multiplicative wealth dynamics have non-ergodic wealth changes, which means that the expectation value of changes in wealth no longer reflects time-average growth. Indeed, there are gambles in which changes in wealth have a positive expectation value, but have a negative time average growth rate[4]. A simple example is a fair coin gamble: heads to gain 50% of one's current wealth, tails to lose 40% of one's current wealth. Counterintuitively, whilst this gamble has a positive expectation value (1.05 times current wealth, per trial), it has a negative time average growth rate (~0.95 times current wealth, per trial). For such gambles, maximising expected value eventually leads to ruin. In such multiplicative settings a logarithmic utility function is time optimal, since maximizing changes in expected utility per unit time then maximises the time average growth rate of wealth[5] (Fig. 1g).

These examples highlight the fact that time optimal behavior relies on agents adapting their utility functions according to the dynamics of their environments. Time optimality here refers to the optimality of a behavioral strategy in maximising the time average growth rate of wealth. A strategy or utility function that affords the maximisation of time average growth of wealth is thus said to be time optimal[§]. In contrast, prevailing formulations of utility theory, including expected utility theory[6,7,8] and prospect theory[9,10,11], are not premised on the dynamics of the environment. In treating all possible dynamics as the same, these formulations imply that utility functions are indifferent to the dynamics. Since standard decision theories assume stable but idiosyncratic utility functions, whereas time optimality prescribes specific utility functions for specific dynamics, the two classes of theory make different predictions. Here we manipulated the ergodic properties of a simple gambling environment, by switching between gambling for additive increments of money versus gambling for multiplicative growth factors, evaluating the effect this has on the utility functions that best account for choices. We found evidence that gamble dynamics impose a consistent effect on utility functions, and that these effects are better approximated by a time optimal model compared to standard utility models.

**Results**

**Methods summary.** We asked whether switching between additive and multiplicative gamble dynamics systematically influences decision making under risk. Specifically our objective is to investigate how existing utility models, primarily prospect theory and isoelastic utility[12], perform in comparison to a null model of time optimality in explaining choice behaviour under changing dynamics. Each subject, in an experiment that spanned two days, engaged in a gambling paradigm with either additive or multiplicative wealth dynamics. At the start of each day, participants were endowed with an initial wealth of 1000DKK / ~$155 (Fig. 1a), after which they took part in a

---

[§] This is distinct from other notions of optimality which typically pertain to the consistency of choices.

passive session during which they had an opportunity to learn, via observation, the deterministic effect of image stimuli on their endowed wealth (Fig. 1b). On the additive day (Day$^+$) the stimuli caused additive changes in wealth whereas on the multiplicative day (Day$^\times$) the stimuli caused multiplicative changes to their endowed wealth (eqs. 1-5, Supp. Fig. 1). Different stimuli were used for the two different days and the association between the stimuli and the change in wealth was randomized between subjects. Having repeatedly observed these contingencies between the stimuli and the changes in wealth, subjects subsequently engaged in an active session during which they chose between two gambles composed of pairs drawn from the same set of stimuli (Fig. 1c, eqs. 6-9). Upon choosing a gamble, each of the two stimuli had a 50% probability of being the outcome of the gamble. Subjects understood that the gamble outcomes were not revealed during the game, and that 10 of the outcomes of the chosen gambles would be randomly applied to their wealth at the end of each day for payout. There were four sessions in total per subject, Passive$^\times$ and Active$^\times$ occurring on Day$^\times$; and Passive$^+$ and Active$^+$ occurring on Day$^+$. We adopted three complementary analysis strategies. The first is model-independent in the sense that we tested whether choice frequencies change according to gamble dynamics. The second and third approaches were model-dependent insofar as we formally compare theoretical models of utility in terms of their parameter estimates, and in terms of the predictive adequacy of each utility model.

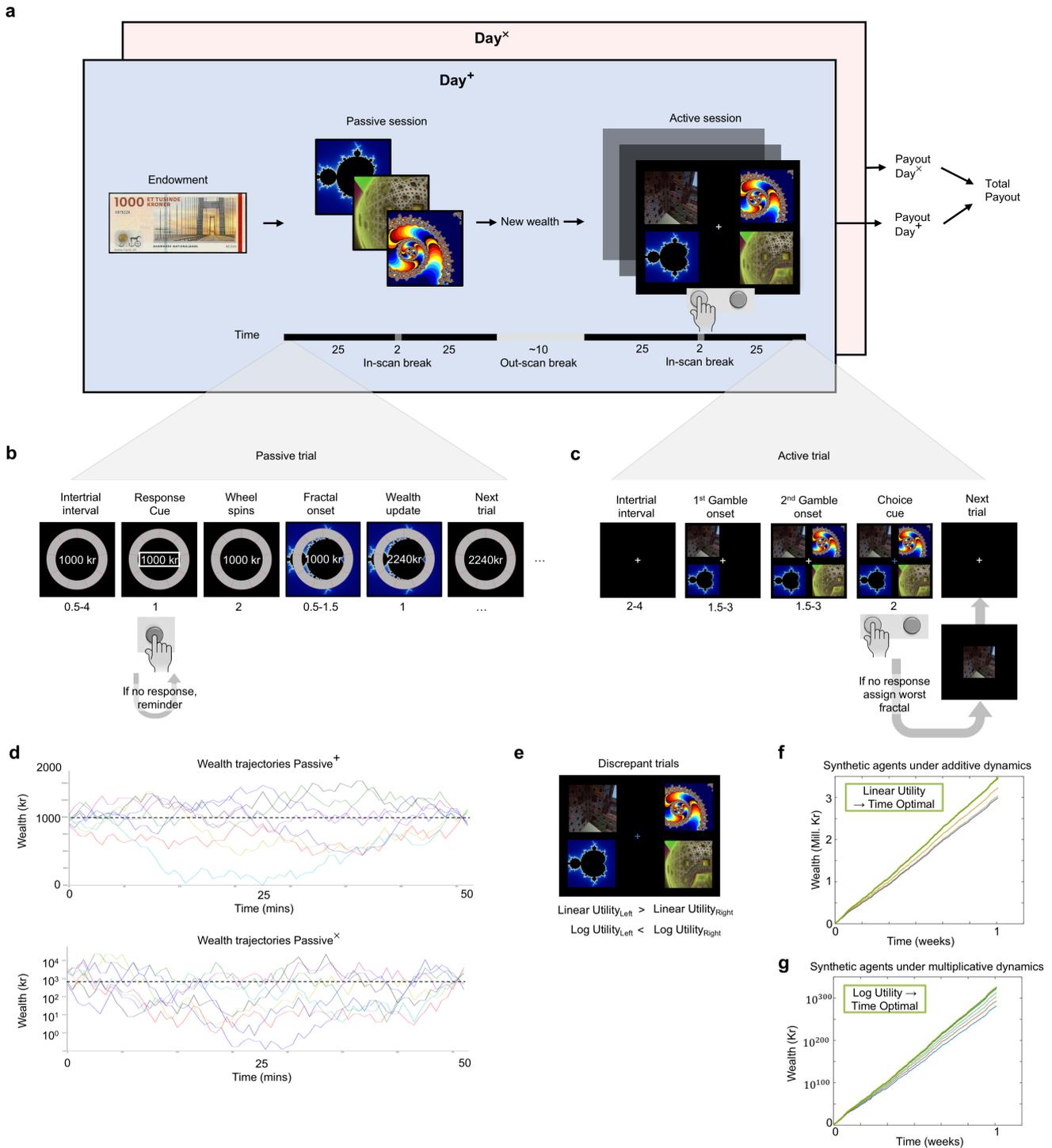

**Figure 1 | Experimental design and wealth trajectories. a**, two-sheets (blue and pink) summarise the repeated protocol for both days, which only differ in the dynamics of wealth changes. Numbers indicate durations in minutes. Although three stimuli are shown for illustration, a total of 9 stimuli were used in each session. **b**, single trial from a passive session, where durations are in seconds and ranges depict a uniformly distributed temporal jitter. **c**, single trial from an active session. **d**, wealth trajectories in real time over the course of each passive session. The trajectory for Passive$^\times$ is plotted on a log scale, appropriate to the multiplicative dynamics. Eight randomly selected trajectories are plotted. Dotted line shows initial endowment level of 1000DKK. **e**, discrepant trials are a subset of trials, where agents with linear and logarithmic utility functions would be predicted to make different choices. In the example here, an agent with linear utility would choose the left-hand gamble whereas an

agent with logarithmic utility would choose the right-hand gamble. **f,** wealth trajectories of synthetic agents with different utility functions (prospect theory and isoelastic) repeatedly playing the set of additive gambles over one week (Supp. Results: Synthetic agents). The agent with linear utility has the highest time average growth rate (green). **g,** equivalent simulations for multiplicative gambles. The agent with log utility has the highest time average growth rate (green). The time optimal agent is an agent with linear utility for additive dynamics, and log utility for multiplicative dynamics, and thus also experiences both of the wealth trajectories depicted in green (in f and g).

**Gamble dynamics affect choice frequencies.** Discrepant trials are the subset of trials in which a linear utility agent would choose a different gamble to a log utility agent (Fig.1e), 25 of 312 trials in the active session had this discrepant property. For example, in the Fig. 1e, an agent with linear utility would be more likely to choose the left gamble, whereas an agent with logarithmic utility would be more likely to choose the right. By observing the choice proportions (CP) we obtain evidence about the dependency between choices and gamble dynamics (Fig. 2a). We quantify evidence in terms of Bayes factors which are defined as the relative likelihood for one model over another, given the observation of the data. A Bayes factor of 10 for model 1 over model 2 indicates that the data is 10 times more likely given model 1 than given model 2. Levels of evidence are reported according to standard interpretations of Bayes factors (BF)[13,14]; ranging from anecdotal (1-3), moderate (3-10), strong (10-30), very strong (30-100) through to extreme (100>). We found moderate evidence against the hypothesis that subjects choose in favour of linear utility ($CP_{log}<0.5$) under additive dynamics (Fig. 2b-d, $BF_{0-}$ = 3.678, $M_{(CP)}$ = 0.4932, SD = 0.1969, SEM = 0.04641, Bayesian central credibility interval: $BCI_{95\%}$ [0.395, 0.591], robust over prior widths). In contrast, we found extreme evidence for the hypothesis that subjects choose in favour of log utility ($CP_{log}>0.5$) under multiplicative dynamics (Fig. 2e-g, $BF_{+0}$ = 460.4, $M_{CP}$ = 0.718, SD = 0.188, SEM = 0.044, $BCI_{95\%}$ [0.625, 0.812], robust over prior widths). Note that choosing in favour of linear utility ($CP_{lin}>0.5$) is equivalent to choosing against logarithmic utility ($CP_{log}<0.5$), and vice versa. The variable choice proportion for the multiplicative condition may not be normally distributed (Shapiro-Wilk p=0.019), and thus we repeat the analysis with an equivalent non-parametric test (Wilcoxon Signed-ranks, V = 159, p < .001, effect size 0.86). Correspondingly, we found very strong evidence for the hypothesis of a within-subject increase in the choice proportions in favour of log utility when dynamics shift from additive to multiplicative (Fig. 2h-k, $BF_{+0}$ = 52.38, $M_{\Delta CP}$ = 0.225, SD = 0.253, SEM = 0.060, $BCI_{95\%}$[0.099, 0.351], robust over prior widths). Finally, averaging across all models that entail possible combinations of factors and covariates, we found that the inclusion of the dynamic as a factor was uniquely favoured by the data (rmANOVA, $BF_{inclusion}$ = 80.2) with all other factors including order of testing showing $BF_{inclusion}$ < 1, see Supp. Fig. 2b). Together, this shows strong evidence that in the discrepant trials, gamble dynamics exert a strong and systematic influence over choices.

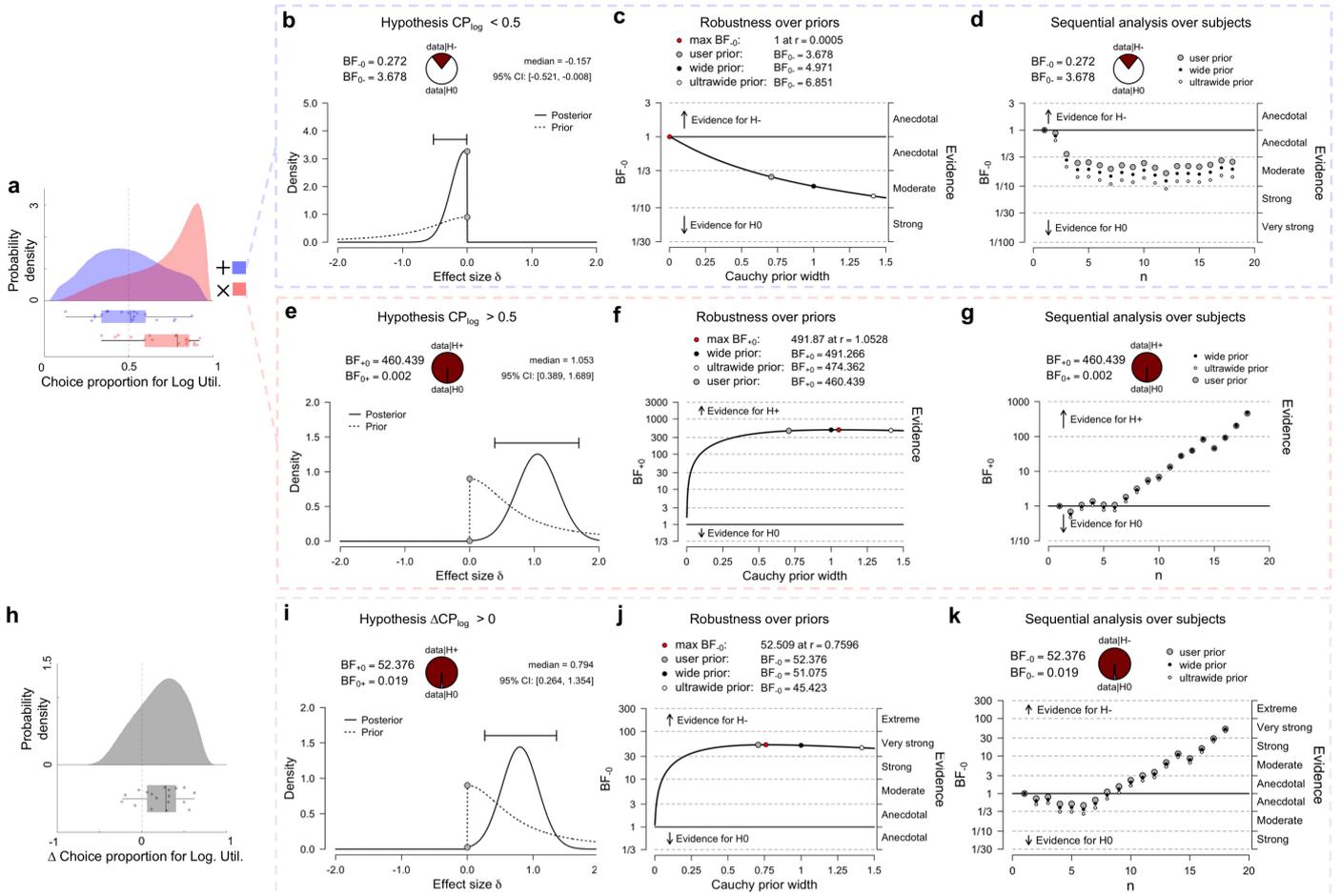

**Figure 2 | Gamble dynamics affect choice frequencies. a,** raincloud plot[15] showing choice proportions in favour of log utility ($CP_{log}$), for multiplicative (red) and additive (blue) dynamic with split-half violin plot (top) and raw jittered data of individual subjects' choice proportions together with box and whisker plot (bottom). All box and whisker plots indicate range, 1st & 3rd quartiles, and median. **b,** prior and posterior density for the hypothesis that choice probabilities are in favour of linear utility ($CP_{log} < 0.5$) in terms of effect size, for the additive dynamic (Bayesian t-test), reporting Bayes factor in favour of $CP_{log}$ being lower than 0.5 (negative effect size, indicated by $BF_{-0}$) and its reciprocal in favour of the null hypothesis ($BF_{0-}$) **c,** robustness analysis of Bayes factors in b, showing that a less informative prior (ultrawide) would increase the Bayes factor in favour of the null hypothesis. **d,** Sequential analysis showing how this Bayes factor changes with increasing numbers of subjects, with the different markers indicating different prior widths. **e-g**, equivalent analyses for the multiplicative dynamic for the hypothesis that choice probabilities are in favour of log utility ($CP_{log} > 0.5$). **h,** raincloud plot of the individual change in choice proportion ($\Delta CP_{log}$) where positive numbers indicate an increase under multiplicative dynamics. **i,** posterior and prior densities for the hypothesis that $CP_{log}$ is larger for multiplicative compared to additive dynamics (Bayesian Paired t-test). **j-k,** equivalent robustness and sequential analyses for this test.

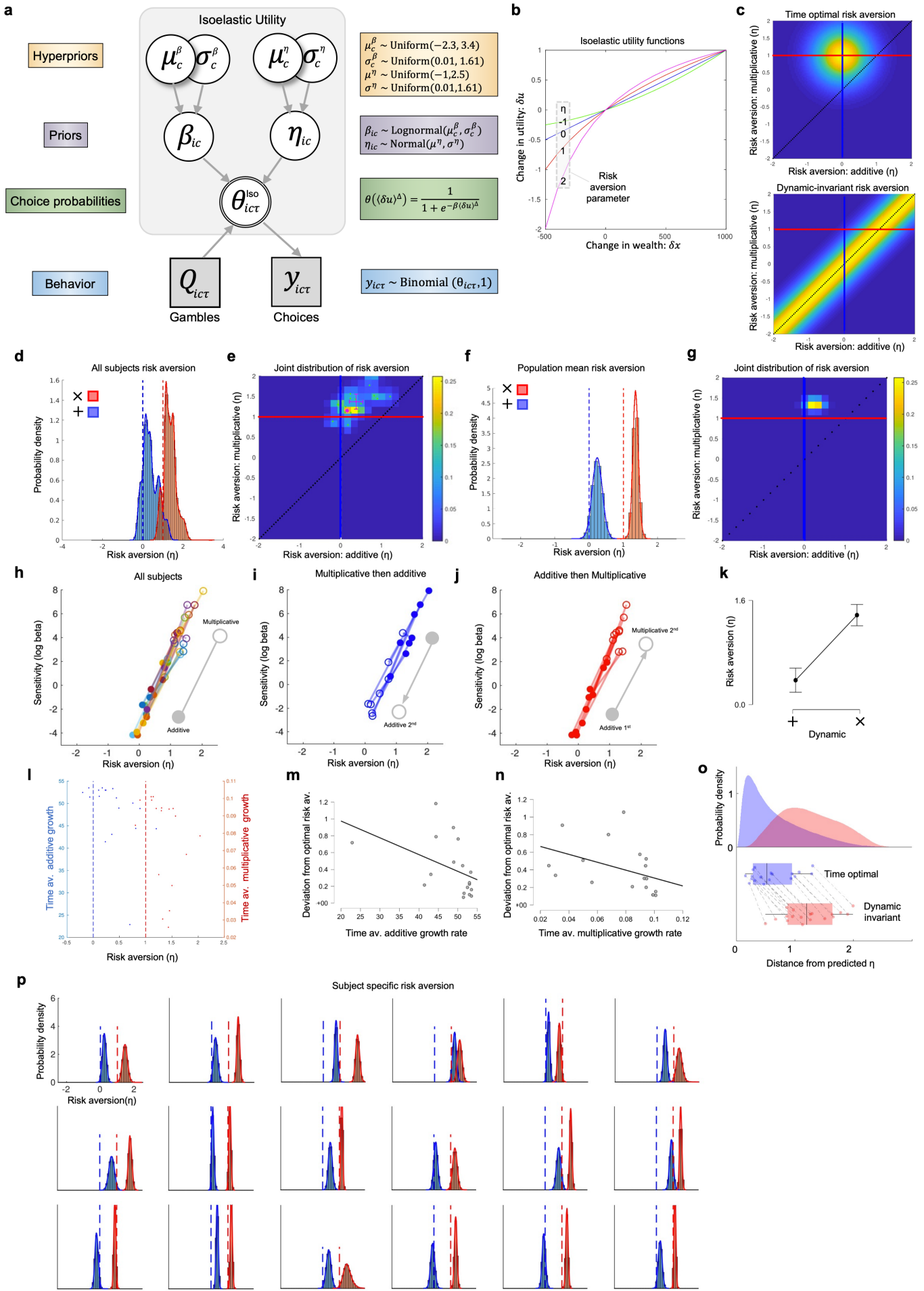

**Figure 3 | Hierarchical Bayesian model for estimating dynamic-specific risk preference. a,** hierarchical Bayesian model for estimating risk preferences. Circular nodes denote continuous variables, square nodes discrete variables; shaded nodes denote observed variables, unshaded nodes unobserved variables; single bordered nodes denote deterministic variables, double bordered nodes stochastic variables. Along the left-hand side describes what role these variables play, and along the right side includes details on the distributions and logistic choice function. The data generating process (blue) which maps from theta to binary choice, is equivalent to a Bernoulli distribution. **b,** spectrum of utility functions entailed by different values of the risk aversion parameter $\eta$. **c,** schematic of model predictions of what values $\eta$ will take for a time optimal (top) and dynamic-invariant isoelastic models (bottom). Heatmaps indicate probability density, with red & blue lines indicating time optimal risk aversion for additive and multiplicative conditions, respectively, intersecting at the time optimal strategy for both dynamics. Diagonal line indicates risk aversions that are invariant to dynamics. **d,** frequency distribution of risk aversion values collapsed over subjects for additive (blue) and multiplicative (red) dynamics. Dotted lines indicate time optimal values of risk aversion. **e,** joint distribution of dynamic-specific risk preferences. Maximum a posteriori (MAP) values are plotted for the group (pink dot), and for each subject (cyan dots), and are superimposed over the group-level frequency distribution. Error-bars indicate the central BCI$_{95\%}$ for the subject-specific MAP values. Red, blue, and diagonal lines have same meaning as in panel c. **f-g,** same as d-e, but for posterior distributions of the population-level mean risk aversion. **h,** displacement in parameter space caused by changing the gamble dynamic. Filled and empty circles indicate additive and multiplicative dynamics, respectively. **i-j,** equivalent displacements splitting subjects according to the temporal order of their experience of the dynamics. **k,** mean risk aversion under each dynamic, bars show central BCI$_{95\%}$. **l,** distribution of subject specific time average growth rates and risk aversion under both dynamics. **m,** correlation between time average additive growth rate of subject's choices and deviation of subject's risk aversion away from the time optimal value. **n,** equivalent plot for multiplicative dynamics. **o,** raincloud plots of Euclidean distances of MAP$_\eta$ estimates to the predictions of the time optimal and dynamic invariant utility models. Grey lines link estimates from the same subjects. **p,** individual posterior probability distributions for risk aversion. Red and blue lines indicate time optimal values for additive and multiplicative dynamics, respectively.

**Estimates for utility model approximate time optimality.** The model-free analysis of choice behaviour in the discrepant trials (25 per subject and condition) suggested that gamble dynamics affect choice behaviour in the direction predicted by time-optimality (Fig. 2). As a next step, we fit an isoelastic utility model (also called constant relative risk-aversion utility function, CRRA) to the entire sample of choices (312 per subject and condition). For a discussion of testing the predictions of multiperiod utility as alternative models see Supplementary Discussion. The isoelastic utility model has a single risk aversion parameter ($\eta$), negative values of which entail risk seeking, zero entails risk neutrality, and positive values entail risk aversion (Fig. 3b. eq. 11). This model is suited to an explorative analysis of time optimality insofar as its parameter space contains values that are time optimal solutions for both additive and multiplicative dynamics. Specifically, an agent that switches from risk neutrality with an $\eta$ of 0 under additive dynamics, and to risk aversion with an $\eta$ of 1 under multiplicative dynamics, is achieving time optimality by switching between linear and logarithmic utility. Thus, from this perspective, risk aversion should be calibrated to the dynamical setting to maximise the time average growth rate of wealth. Such time optimal agents would be expected to distribute their $\eta$ parameters around this optimal point as in Fig. 3c (upper panel), whereas agents with no systematic shift (dynamic-invariant agents), would distribute around the diagonal line (lower panel). In estimating a hierarchical Bayesian model of isoelastic utility (Fig. 3a), we obtained separate posterior distributions of risk aversions for each gamble dynamic, which can be compared to these theoretical predictions. We refer to this as a dynamic-specific isoelastic model. Firstly, we find extreme evidence that risk aversion increases from additive to multiplicative dynamics (Fig. 3k, Paired-t, $BF_{10}$ = 2.9 × $10^7$, $M_\Delta$ =1.001, SD = 0.345, SE = 0.081, $BCI_{95\%}$[0.829,1.172]), which is indistinguishable from the predicted size of change in $\eta$, under time optimality. As with the choice proportions, we found extreme evidence for the effect of gamble dynamic on risk aversion, compared to all other factors tested (rmANOVA $BF_{inclusion}$ = 2.45 × $10^9$, all other factors < 1, Supp. Fig. 4b). The same effect was evident when including covariates that account for differences in variance in wealth and wealth changes during the passive phase (rmANOVA $BF_{inclusion}$ = 1.19 x $10^{10}$, all variance factors < 1, Supp. Fig. 4f). Finally the frequency histograms of risk aversion marginalised over all subjects (Fig. 3c) show that the maximum a posteriori (i.e. the most likely value of the posterior parameter distribution, $MAP_\eta$) value approximates the time optimal predictions for each dynamic: under additive dynamics, the distribution estimated from the data has a $MAP_\eta$= 0.1506, compared to the time optimal prediction of $\eta$ = 0 (Fig. 3d, blue); under multiplicative dynamics, the distribution estimated from the data has a $MAP_\eta$=1.1534, compared to the time optimal prediction of $\eta$ = 1 (Fig. 3d, red). The joint distribution over a risk aversion space (Fig. 3e) shows that the MAP estimate of the joint distribution is likewise close to the optimal point indicated by the intersection of the prediction lines. A complementary visualisation of this correspondence comes from the posterior distribution of the population parameter for the mean of $\eta$ (Fig. 3f-g). This indicates a qualitative agreement between the distribution of risk aversions, and the normative predictions of the time optimality model.

**Risk preferences are closer to predictions of time optimality.** To test whether risk aversion values are explained better by time optimality (Fig. 3c upper), or alternatively by a dynamic invariant utility model (Fig. 3c, lower), we computed the distance of each subject's risk aversion ($MAP_\eta$) to the predictions of each model. For the time optimal model this is the Euclidean distance to the time-optimal coordinate (0,1), and for the dynamic invariant model this is the distance to the closest point on the diagonal. We find extreme evidence that risk aversions are closer to the time optimal prediction (Fig. 3o, Paired-t, $BF_{10}$ = 2.8 × $10^{11}$, M = 0.623, $BCI_{95\%}$ [0.565, 0.681], Supp. Fig. 3e-h), and that this is true for every subject tested. Together this shows that the time optimality model is a better predictor of risk aversion over different dynamics, than a null model which assumes no effect of dynamics on risk aversion.

**Order of gamble dynamics does not substantially affect choice.** In the dynamic-specific isoelastic model, both the risk aversion parameter $\eta$ and the sensitivity parameter $\beta$ (modelling how sensitive choices are on differences in utility, eq. 15) are free to vary for each subject when the gamble dynamics change (Fig. 3a). Plotting the joint distribution of both $\eta$ and $\beta$, affords visualisation of the effect of the dynamic on both risk aversion and on choice sensitivity (Fig. 3h). We found that a switch from additive to multiplicative dynamics is associated with a characteristic shift in this parameter space toward greater risk aversion, and toward greater sensitivity. The order in which subjects experienced different gamble dynamics was counterbalanced over subjects. In the subgroup that tested in the additive condition first (Fig. 3j), the movement in parameter space is in the opposite direction to the subjects tested multiplicative condition first (Fig. 3i), as predicted if the effect was primarily driven by the dynamic and not the order of testing. The inclusion probability for the order of testing had a Bayes Factor below one, indicating anecdotal evidence that the data disfavours its inclusion in the model ($BF_{inclusion}$ = 0.891, Supp. Fig. 4c). Thus, there is no statistical evidence that the order of exposure to different gamble dynamics substantially affected choice.

**Deviation from time optimal value decreases time average growth rates for wealth**. The relation between a subject's risk aversion and the time average growth rate of their choices (eqs. 8-9) can be noisy due to the probabilistic relation between utility and choice. This stochasticity is visible in the relation between the time average growth rates of the choices made and the risk aversion estimated for each subject under both dynamics, though the highest growth rates coincide with values close to the time optimal risk aversion (Fig. 3l). Further, we found that the closer the subjects shifted their risk aversion toward time optimal values, the higher the time average growth rates of their wealth, given their choices for both additive (Fig. 3m, τ = -0.428, $BF_{10}$ = 10.51, $BCI_{95\%}$ [-0.655, -0.086]) and multiplicative dynamics (Fig 3n, τ = -0.502, $BF_{10}$ = 30.88, $BCI_{95\%}$ [-0.711, -0.131]). Thus, the risk aversion parameter that best describes a subject's choices is predictive of their time average growth rate. This illustrates that deviating from time optimality has negative consequences for growing wealth, as implied by theory.

**Bayesian model selection supports time optimality over other utility models.** The dynamic-specific isoelastic utility model suggested that subjects dynamically adapt their choice behaviour in a way predicted by time optimality. We next compared the predictive adequacy of three models, an isoelastic model, a prospect theory model (eq. 10), and the time optimal model (eq. 12), detailed in Fig. 4a&b. The time optimal model is fixed in its theoretical predictions for the population means of $\eta$, restricted to be 0 for additive dynamics and 1 for multiplicative dynamics. However, the variance around this mean is a free parameter in order to account for the plausible assumption that not all subjects are phenotypically identical. Prospect theory has two utility parameters whose means are not fixed at the population level but are free to vary within standard restrictions that define the theory (See Models, in Methods section). Finally, the isoelastic model has one utility parameter that is estimated across both sessions, whose mean is free to vary at the population level. Markov chain Monte Carlo sampling of this model results in posterior frequencies for the model indicator variable $z$ that are interpreted as posterior probabilities for each model, estimated for each subject[16]. Most subjects had most of their probability mass located over the time optimal model (Fig. 4c), as is evident from the marginal probability over subjects (Fig. 4d). Computing protected exceedance probabilities, which measure how likely it is that any given model is more frequent (estimated frequencies in Fig. 4e) than all other models in the comparison set, we found that the time optimal model had an exceedance probability of 0.976 (Fig. 4f) which corresponds to very strong evidence for being the most frequent (BF$_{\text{Time-PT}}$ = 76.9, BF$_{\text{Time-Iso}}$ 80.6).

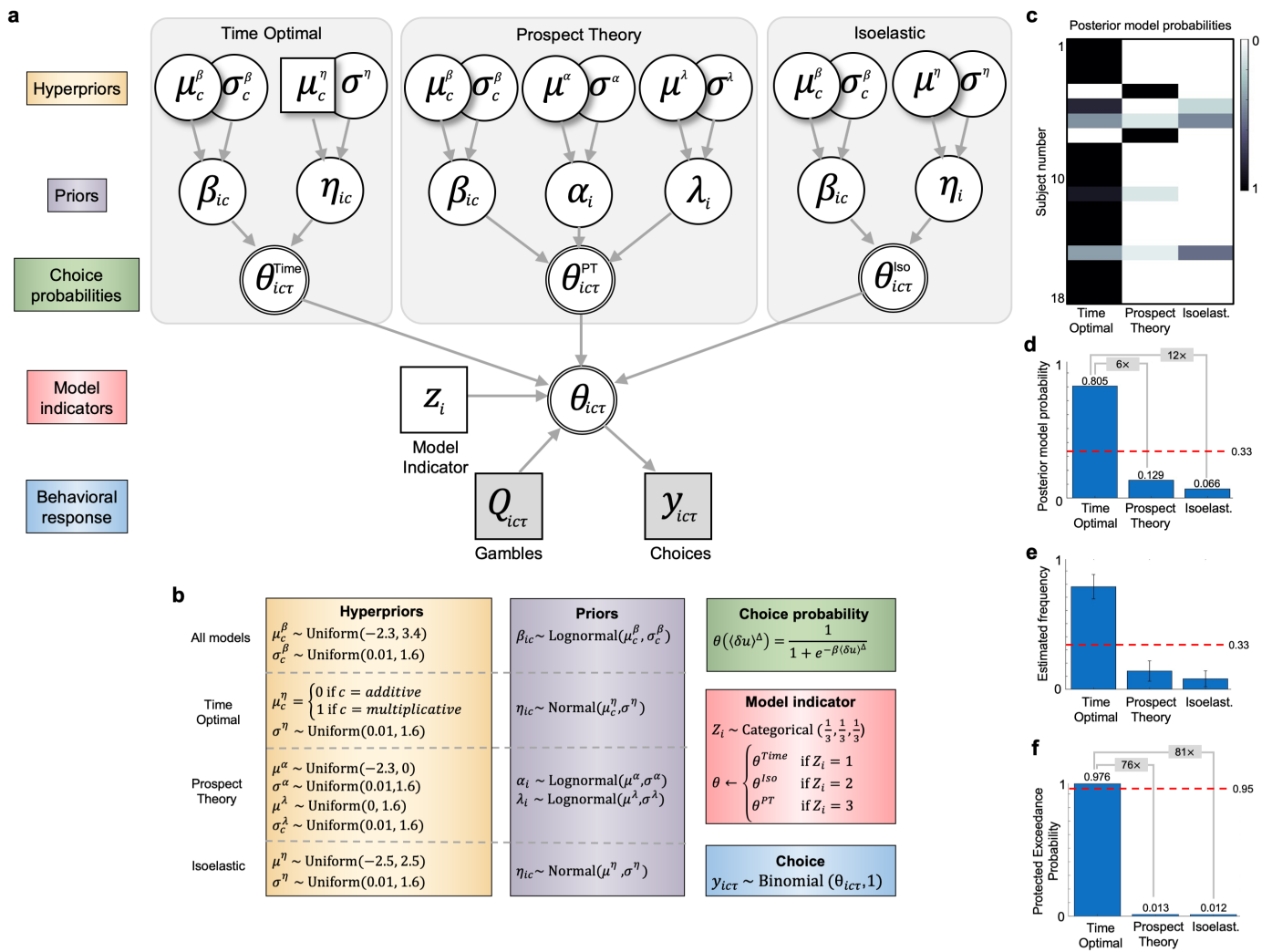

**Figure 4 | Bayesian hierarchical latent mixture model and model selection results. a,** graphical model according to conventions of Fig. 3. This model adds a model indicator variable (z) to modelling latent mixtures of the three different utility models nested within it. Note that for prospect theory risk preference parameter $\alpha$, there is one parameter for gains, and another for losses. **b,** hyperprior and prior distributions, including structural equations, choice functions, and choice generating distributions. Hyperpriors for $\alpha$ are duplicated to model gains and losses separately. **c,** posterior model probabilities for each model based on the model indicator variables representing each utility model. **d,** posterior model probabilities summed over subjects, with the red bar indicating prior probabilities assuming equal prior probability for the three utility models. **e,** estimated model frequencies from the cohort and error bars as standard deviations. **f,** protected exceedance probabilities for each utility model being the most frequent.

## Discussion

**Summary.** By manipulating the dynamical properties of simple gambles, we show that ergodicity-breaking can exert strong and systematic effects on risk-taking behavior. Switching from additive to multiplicative gamble dynamics reliably increased risk aversion, which in most subjects tracked close to the levels that maximise the time average growth of wealth. We show that these effects are well approximated by a model of time optimality based on ergodic theory and cannot be adequately explained by the prevailing models of utility in economics and psychology.

**Main findings.** The time optimal model assumes that agents prefer their wealth to grow faster, and that this preference for faster growth is stable. From these assumptions, it can be shown that to maximise the time average growth rate of wealth, agents should adapt their utility functions according to the wealth dynamics they face, such that changes in utility are rendered ergodic[5]. From this, a number of simple predictions can be derived, each increasing in specificity. First, to approximate time optimal behavior, different gamble dynamics require different ergodicity mappings. Thus, when an agent faces a different dynamic, this should evoke the observation of a different utility function. This was observed, in that all subjects showed substantial changes in their estimated utility functions (Fig. 3p). Second, in shifting from additive to multiplicative dynamics, agents should become more risk averse. This was also observed in all subjects. Third, the predicted increase in risk aversion should be, in the dimensionless units of relative risk aversion, a step change of +1. The mean step change observed across the group was +1.001 (BCI$_{95\%}$ [0.829,1.172]). Fourth, to a first approximation, most (not all) participants modulated their utility functions from ~linear utility under additive dynamics, to ~logarithmic utility under multiplicative dynamics (Fig. 3d). Each of these utility functions are provably optimal for growing wealth under the dynamical setting they adapted to[5], and in this sense they are reflective of an approximation to time optimal behavior. Finally, model comparison revealed strong evidence for the time optimal model compared to both prospect theory and isoelastic utility models, respectively. The latter two models provide no explanation or prediction for how risk preferences should change when gamble dynamics change, and even formally preclude the possibility of maximising the time average growth rate when gamble dynamics do change. In line with this explanatory gap, both prospect theory and isoelastic utility models were inadequate in predicting the choices of most participants (Fig. 4c).

**Differences between conditions.** In the passive phase of the experiment, in which subjects learnt the effects of the stimuli on their wealth, there were different dynamics at play in the different conditions that could in principle lead to differences in the experience of the subject. One such difference is in the variance of the wealth changes, which were higher on average for the multiplicative than the additive condition. These variances were variable across subjects, however we found no evidence for them explaining the observed differences in risk aversion (Supp Results, Supp. Fig. 4f). It is possible that other features of the wealth trajectories may differ between the two conditions, which is unavoidable due to the two dynamics being qualitatively different. One such difference is due to the fact that since wealth levels needed to be bounded between 0 and 5000kr at all times during the passive phase, many of the possible wealth trajectories had to be discarded when sampled during the experimental setup, prior to each subject's session. For the multiplicative condition, substantially more positive excursions were discarded, whereas for the additive conditions more negative excursions were discarded. While this skews the representativeness of the random process generating the stimuli and makes it non-independent, it should be noted that this is irrelevant to the decision making phase which are drawn from a qualitatively different generative process. Furthermore, the selective filtering imposed by this process acted to attenuate the differences in the wealth trajectories between conditions. Another consideration is whether subjects learnt the stimuli better under the additive condition. Such an effect should be apparent in the distribution of risk aversion parameters, in which greater uncertainty should manifest in less precise posterior distributions. This was not observed (Fig. 3p). To date, there has been one partial replication of this study, focusing on the fidelity with which subjects can discriminate these stimuli, finding that there is no strong evidence for a difference between conditions, even when stimuli are learnt in one fifth of the time[21].

**Statistical considerations.** The size of the cohort (achieved n=18) was constrained to concentrate power within subjects, and by the high-stakes design, in which each participant could walk away with up to 750 USD in payout. Restricting our inferences to this cohort, the effect was consistent across all participants, and was reproducible across different inferential approaches. In general, the strength of the evidence we obtained from individuals likely derives from the fact that the game is high stakes, and also from the fact that we collected a large number of decisions (over 600 per participant) over a large number of distinct gambles (320 per participant). This affords opportunity for stringent testing between utility models that make overlapping predictions. The strength of the evidence thus observed, likely derives from this being a large and consistent effect size, that was likely driven by the large incentives, a fundamental shift in strategy caused by the dynamics, and by the large number of trials. Indeed, for many of the tests conducted, high degrees of evidence are reached before reaching the full subject group. Finally, the fact that discrimination between utility models is possible under our modelling framework is evident from its ability to recover parameters and model identities from synthetically generated agents (Supp. Fig. 5).

**Validity.** The ethical constraint of not allowing subjects to lose money at the end of the experiment potentially impacts more on the additive condition, since negative wealth is impossible under multiplication of positive growth factors. Strictly speaking, the prediction that linear utility is the time optimal utility function for additive dynamics assumes only additive dynamics without any such constraints. The fact that the data are reasonably well explained by a theory which ignores these constraints suggests that, to a first approximation, these constraints are not critical for predicting the behavior of these participants. We are careful not to make formal claims about the generalisation of this time optimality behavior, beyond the subjects tested, and beyond the paradigm used. Establishing time optimality as a general phenomenon will require multi-centre replication, and then broader generalisation to ascertain its robustness to paradigm variations.

**Theoretical considerations.** The dependency between dynamics and risk aversion that we observe here is relevant to a widespread assumption that utility functions are stable over time[17–19,23]. Primarily, this is motivated on epistemological grounds. If utility is to predict behaviour in future settings, then it must be stable, otherwise if behavior changes, it is not known if this is due to a change of setting or preference, or both[16,26]. However, this is contradicted by multiperiod utility models[27], as well as a diversity of empirical demonstrations of preference instability. In animals, including humans, there is evidence suggesting that risk preferences depend on homeostatic [,29,30,31], circadian[32], and affective states[33]. Test-retest stability in the same settings, though typically reported as modest[34], can be relatively high when estimated using hierarchical models of the sort used here[35]. The findings reported here place the stability of utility in a broader context by connecting to an optimality framework for how utility functions should change in response to changes in one's environmental dynamics. This casts the dynamical dependence of utility functions observed here, not as preference instability per se, but simply as a manifestation of a stable preference for growing wealth over time when facing different circumstances.

**Final remarks.** Models of decision-making are predominantly developed without recourse to dynamical considerations and are typically tested in settings that implicitly evoke additive dynamics. The theories developed under these conditions are then assumed to generalise to settings in which additive dynamics may no longer apply, and multiplicative dynamics likely dominate, which may contribute to the predictive inadequacy of many models. In light of this initial evidence, we suggest this motivates a need to develop and further scrutinise theories of decision making that are explicitly conditioned on ergodic foundations.

## Methods

**Subjects, Power, Ethics.** This paper focuses on the behavioral data obtained from a neuroimaging study on the neural encoding of utility. The criteria for inclusion were being aged 18-50, and fluent in English. The criteria for exclusion were a history of: psychiatric or neurological disorder, credit problems (operationalized via bad pay status on www.dininfo.dk), or expertise in a quantitative or cognitive domain (finance, banking, accountancy, economics, mathematical sciences, computer science, engineering, physics, psychology, neuroscience). MRI-specific exclusion criteria were also applied, including implanted metallic or electronic objects, heart or brain surgery, severe claustrophobia, or inability to fit into the scanner (weight limit of ~150kg, bore diameter of 60 cm). Except for the latter, all such information was self-reported. The intended sample size was 20, however due to post-hoc exclusion (1 participant fell asleep, 1 failed to learn the stimuli) the achieved sample size was 18 (6 females, age: M = 25.79, SD = 4.69, range 20-38). Subjects were recruited as a convenience sample, via the subject recruitment website www.forsøgsperson.dk. The sample number was based on general guidelines for the minimal number of subjects required for medium effect sizes in neuroimaging datasets[36]. The number, timing, and jittering (randomised timing) of events within each session was based on prior efficiency simulations for similar neuroimaging paradigms. As such, no a priori design analyses were performed for the behavioral data only. No stopping rule or interim analyses were performed. Data collection ran from the 10/06/2017 to 30/07/2017. All data was acquired at the Danish Research Centre for Magnetic Resonance. Informed consent was obtained from all subjects as approved by the Regional Ethics Committee of Region Hovedstaden (protocol H-17006970) and in accordance with the declaration of Helsinki. Independent of their payouts in the gambling paradigm, all subjects were compensated 1020 DKK / ~$160 for a grand total of 6 hours of participation over the two days. A forthcoming paper will focus primarily on the neuroimaging data.

**Experimental procedure.** After changing into hospital gowns subjects were read the instruction sheet (see below and Supp. materials). To précis, subjects were truthfully informed that the aim of the experiment was to study how the brain reacts to changes in wealth, that all of the money involved is real, and that the total accumulated wealth will be paid out as the sum of that accumulated over the two days (Fig. 1a). They then played ~20 demo trials of the paradigm in the scanner control room, including both active and passive sessions (~5mins) for no financial consequence. The experimenter demonstrated what happened if buttons were not pressed in time (Fig. 1b&c). Subjects were instructed that each day lasts 3 hours in total, with ~60mins for the passive session (inc. time for localiser scan and shim), a short break, then ~60-75mins for the active session (inc. localiser, shim, anatomical scans), with short breaks within the session (Fig. 1a). Each subject entered the scanner, was set up with a respiratory belt to monitor breathing, and with a pulse meter on the middle or index finger of the non-responding hand. All stimuli were projected under dark conditions onto a screen located within the bore of the MRI magnet (Siemens, MAGNETOM Prisma), and viewed via mirrors mounted to the head coil. Subjects were instructed to fixate the central fixation cross at all times (Fig. 1b&c) and choose via button box. The paradigm was presented via the Psychopy2 toolbox (v1,84.2) running on Python (2.7.11).

**Experimental design.** The experiment is a fully crossed randomized controlled trial in which the wealth dynamic is the primary independent variable, and choice is the primary observable. The wealth dynamic, as well as the deterministic association between stimuli and outcomes was controlled via computer programme and thus double blinded. Further, since payouts at the end of the test day were subject to being randomly realized from each subjects' choices as well as being statistically balanced between conditions, payout was also effectively double-blinded. Subjects were neither informed of any explicit details concerning dynamics or differences between test days, nor given any reason to expect that the test days were different. The instructions, procedures and setup were otherwise identical for both test days. The order in which multiplicative and additive test days were conducted was counterbalanced across the group. Subjects were not able to make notes or use a calculator due to their location inside the brain scanner. The primary measures were the choices acquired during the active session. Measures collected but not included in this report include all functional and structural neuroimaging modalities, physiological noise measurements (pulse rate and breathing), and reaction times. To ensure good quality model estimation, we recorded a large number of decisions (312 in total per active session) spanning a large subspace (144) of the possible unique gamble combinations. To avoid the problems associated with gambling for "peanuts"[37], the outcomes of decisions are for large quantities of money on each trial (mean possible change in wealth Day$^\times$ = 413.07 DKK / per decision, SD = 249.78, range = -422.87 to 946.71, mean possible change in wealth Day$^+$ = 267.76 DKK per decision, SD = 119.20, range = -428 - 428). Subjects were thus strongly incentivized to pay attention to all the stimuli and to optimize their decision-making throughout the active sessions.

**Pre-registration and deviations.** The experimental protocol was preregistered at www.osf.io/9yhau. There was one deviation from the protocol: The preregistration stated that in the Passive$^+$ session, the final additional stimulus applied to their wealth after having returned to 1000DKK (see section "Passive session stimulus sequences" below) would exclude the most extreme stimuli. Those were, however, included in the paradigm.

**Passive session instructions.** Subjects were instructed in English as follows: *"For the passive phase, you will see a number in the middle of the screen, this is your current wealth for the day in kr. When you see a white box around the number, you are to press the button within 1s. (If you do not, you will be instructed to "press button earlier"). Shortly after pressing the button you will see an image in the background, and this will cause your wealth to change. You are instructed to attend to any relationship between the images and the effect this has on your wealth, since in the active phase that follows you will be given the opportunity to choose images to influence your wealth. Learning these relationships can make a large difference to your earnings in the active phase."* These instructions were identical on both days in order to not bias the subject toward any particular strategy.

**Passive session dynamics.** Formally the passive session can be described as follows: At the start of each test day, subjects were endowed with an initial wealth $x(t_0)$ of 1000DKK, which defined their wealth at the first timepoint, which we denote as $t_0$. Independently for each subject, 9 stimuli were randomly assigned (from a fixed set of 18) for Day⁺, with the remaining 9 assigned to Day˟. Each stimulus, viewed at time $t$ was programmed to have a deterministic effect on the subject's wealth $x(t)$, with the sequence of stimuli causing stochastic fluctuations in wealth (Fig. 1d). The sequence of stimuli deterministically caused dynamics in their wealth which can be expressed as:

$$x(t + \delta t) = x(t) \circledast s(t), \qquad (eq.1)$$

where $\circledast$ is a wildcard operator, which on Day⁺ is the addition operator +, and on Day˟ is the multiplication operator ×. $s(t)$ is a random outcome variable drawn from set $S^\times$ on Day˟, and from set $S^+$ on Day⁺ (see Supp. Fig. 1a). This means that the type of wealth dynamic that the stimuli caused was controlled by the test day. On Day˟ under multiplicative dynamics, the outcome $s(t)$ is the realisation of a random multiplier (growth factors) that can range from ~doubling at one extreme, to ~halving at the other (equally spaced on a logarithmic scale). On Day⁺, under additive dynamics, the outcomes $s(t)$ is the realisation of a random increment, ranging from +428 to -428DKK (equally spaced on a linear scale). Though the dynamics are qualitatively different, we set the bounds of the random increments for Passive⁺ to the central 85th percentile interval of the absolute wealth changes on Day˟.

**Passive session stimulus sequences.** The stimulus sequence was randomized such that wealth levels were constrained to lie in the interval $(0\ kr, 5000\ kr)$ at all times. This was achieved by presenting each of the 9 stimuli 37 times (and the ensuing effect on wealth, thus generating a set of 333 stimuli. The sequence order was randomised without replacement. Any sequence that resulted in a partial sum larger than 5000 or lower than 0DKK, would be rejected and another random sequence generated. This was necessary to render the experiment subjectively plausible, and to avoid debts, which for ethical reasons could not be realised. Since each stimulus was presented with equal frequency, at the end of these 333 trials in the additive condition, the finite time average additive growth rate was zero kr per unit time. Equivalently, at the end of the 333 trials in the multiplicative condition, the finite time average multiplicative growth rate amounted to a growth factor of one per unit time. Thus, at the end of these 333 trials, in both conditions subjects had returned to their initial endowed wealth of 1000 DKK. One additional stimulus was then shown and applied to their wealth, meaning that all subjects had a randomly determined wealth level, as they had been informed (Fig 1d).

**Passive session wealth trajectories and growth.** The wealth at the end of the Passive⁺ session can be calculated as:

$$x(t_0 + T\delta t) = x(t_0) + \sum_{\tau=1}^{T} s(\tau), \qquad (eq.2)$$

and for the Passive˟ session as:

$$x(t_0 + T\delta t) = x(t_0) \prod_{\tau=1}^{T} s(\tau), \qquad (eq.3)$$

where, in both equations, $s(\tau)$ is the random outcome variable in round $\tau$, and T is the total number of trials in the passive sessions. The finite time average growth of wealth on Day⁺ can be calculated as:

$$\bar{g}^+_{\Delta t} = \frac{\Delta x}{\Delta t}, \qquad (eq.\,4)$$

where $\Delta x = x(t_0 + T\delta t) - x(t_0)$, and $\Delta t = T\delta t$. On Day$^\times$ this is calculated as:

$$\bar{g}^\times_{\Delta t} = \frac{\Delta \ln x}{\Delta t}. \qquad (eq.\,5)$$

This design ensured substantial opportunity for subjects to learn the causal effects of each stimulus, whilst also not accumulating extremely high or low wealth levels.

**Active session instructions.** After the passive session, the subjects had a short break of ~5mins outside of the scanner before returning to engage in an active choice task in which they repeatedly decided between two different gambles composed of the stimuli they had just learnt about (Fig. 1a). Subjects were instructed as follows: *"With the money accumulated in the passive phase, you will play gambles composed of the same images. In each trial, you will be presented with two of the images that you have learned about in the passive phase. By pressing the buttons in the scanner to move a cursor, you now have the option to choose to either: a) Accept gamble one, in which case you will be assigned one of the two images, each with 50% probability (not shown), or... b) Accept gamble two, in which case you will be assigned one of the two images, each with 50% probability (again not shown). The outcomes of your gambles will be hidden from you, and only 10 of them will be randomly chosen and applied to your current wealth. You will be informed of your new wealth at the end of the active phase. You can keep any money accumulated after the active phase. If you do not choose in time, then we will give you one of the worst images, it is recommended that you always choose in time."* These instructions were identical on both days in order not to bias the subject toward any particular strategy.

**Active session gambles.** As shown in Fig. 1c, within a trial, subjects first saw the first gamble of a pair of gambles. This gamble is composed of two stimuli on the left-hand side of the screen, each of which they knew has a 50% chance of being applied to their wealth should this gamble be chosen. We refer to this as the left gamble, $Q^{(Left)}$. 1.5-3 seconds later (uniformly distributed), on the right they saw another two stimuli, here comprising the right gamble $Q^{(Right)}$. In a two alternative forced choice, on each trial, subjects choose via button press between gamble $Q^{(Left)}$ and $Q^{(Right)}$. Formally the gambles are:

$$Q^{(Left)} = \begin{cases} s_1^{(Left)}, p_1^{(Left)} = 0.5 \\ s_2^{(Left)}, p_2^{(Left)} = 0.5 \end{cases} \qquad (eqs.\,6\&7)$$

$$Q^{(Right)} = \begin{cases} s_1^{(Right)}, p_1^{(Right)} = 0.5 \\ s_2^{(Right)}, p_2^{(Right)} = 0.5 \end{cases}$$

Choosing between two gambles eliminates any confounds caused by potential preferences for or against gambling[38]. Note that all probabilities are equal and correspond to a fair coin, such that these are easily communicated and control for any probability distortion effects. The outcome of each gamble was hidden from subjects to avoid subjects being "conditioned" to prefer particular stimuli as a function of the stochastic pattern of previous outcomes. This also prevents mental accounting, where subjects keep track of what they have earnt, which introduces idiosyncratic path dependencies.

**Active session growth rates.** For any gamble we can calculate its time average growth rate. The time average additive growth rate for the left-hand gamble is:

$$\bar{g}^{+(Left)} = \left\langle \frac{s^{(Left)}}{\delta t} \right\rangle, \qquad (eq.8)$$

and equivalently for the right-hand gamble. The time average multiplicative growth rate for the left-hand gamble is:

$$\bar{g}^{\times(Left)} = \left\langle \frac{\ln s^{(Left)}}{\delta t} \right\rangle, \qquad (eq.9)$$

and equivalently for the right-hand gamble. Note that the angled brackets indicate the expectation value operator. Note from that there were no numerical or symbolic cues at this point, their decision could only be based on their memory of each stimulus (Fig. 1c). If subjects did not respond within the decision window, then they were assigned the worst stimulus for that trial.

**Active session gamble space.** For any one gamble, there are 81 possible combinations of stimuli ($9^2$, see Supp. Fig. 1b), and 6561 possible pairs of gambles ($81^2$). This gamble-choice space is too large to exhaustively sample, and contains many gambles that do not discriminate between our hypotheses, and thus we imposed the following constraints: All gambles should be mixed (composed of a gain and a loss), and no two stimuli presented in one trial should be the same, this reduces the gamble choice space down to 144 unique non-dominated choices between gambles - 16 mixed gambles (red text cells, in Supp. Fig. 1b) , paired with 9 other mixed gambles with unique stimuli, gives 16*9 possible gamble pairs. Each of these choices was presented twice, resulting in 288 in total. This restriction of the gamble space thus provides a more efficient means of testing the competing hypotheses of this experiment. Subjects were also presented with 24 No-brainer choices, in which both gambles shared an identical stimulus, but differed in a second. These are otherwise known as statewise dominated choices. In these No-brainer choices, the subject should choose whichever gamble includes the better unique stimulus. This offers a direct means of testing of whether subjects could accurately rank the stimuli. One participant (#5) failed to choose statewise dominated gambles with a probability > 0.5 and was excluded from further analysis (Supp. Fig. 4e). All choices were presented in a random order without replacement.

**Subject payout.** Subjects were informed of the following on the first test day prior to the passive session: *"At the end of the two days. Your accumulated wealth will be added over the two days, and transferred to your account, within approximately two weeks, and is taxable under standard regulations (B-income). Total earnings = (Wealth after day 1) + (Wealth after day 2). This will be paid over and above your remuneration for participating in the experiment."* Payout on each test day was limited to the range of 0 to 2000DKK for each day, and thus the range of possible grand total payouts was 0 to 4000DKK (excluding compensation for time).

# Models

**Model summary.** The aim of the modelling was to perform both parameter estimation and model selection. All models deployed hierarchical Bayesian methods, estimated via Monte Carlo Markov Chain sampling. For parameter estimation we estimated a hierarchical model of isoelastic utility, whereas for model selection we estimated a hierarchical latent mixture model, to model latent mixtures of three different utility models.

**Model space.** Following[11] models can be described by specifying three functions: a utility function, a stochastic choice function, and probability-weighting function. Since all probabilities of outcomes are identical in our experiment, we do not deploy any probability-weighting function. The principal objective of the modelling is to compare between different utility functions in accounting for the choice data over both dynamical conditions. We compared three utility models:

**Prospect theory** where changes in utility are equal to a power function of changes wealth:

$$\delta u = \begin{cases} (\delta x)^{\alpha_{gain}} & if\ \delta x > 0 \\ -\lambda |(\delta x)|^{\alpha_{loss}} & if\ \delta x \leq 0 \end{cases}, \qquad (eq.\,10)$$

where $\alpha_{loss}$ and $\alpha_{gain}$ are risk preference parameters lying on the interval (0,1), and $\lambda$ is a loss aversion parameter which lies on the interval $(1, \infty)$. Note that, although this is referred to as value within prospect theory itself, we here refer to this as utility for clarity of comparison between models.

**Isoelastic utility** where changes in utility are given by:

$$\delta u = \delta x \cdot x^{-\eta}, \qquad (eq.\,11)$$

where $\eta$ is a risk aversion parameter which lies on the real number line, with risk aversion increasing for numbers above 0, and risk seeking increasing for increasingly negative numbers. This is obtained by taking the derivative of the isoelastic utility function with respect to changes in wealth.

**Time optimal utility** where changes in utility are determined by linear utility under additive dynamics, and by logarithmic utility under multiplicative dynamics.

$$\delta u = \begin{cases} \delta x & if\ additive\ dynamics \\ \delta \ln(x) & if\ multiplicative\ dynamics \end{cases}. \qquad (eq.\,12)$$

Note that this model follows from one criterion, that agents maximise the time average growth rate of their wealth according to the dynamic they face. These utility functions allow the time average growth rates under these two dynamics to be computed and maximised by choice.

**Expected utility.** For each gamble the expected utility is calculated for each utility model as the expectation value:

$$\langle \delta u^{Left} \rangle = p \cdot \delta u_1^{Left} + p \cdot \delta u_2^{Left} \qquad (eq.\,13)$$

and equivalently for the right-hand gamble. Differences in utility between the left and right gambles are denoted by $\Delta$ such that the difference in expected utility between the left and right-hand gamble is

$$\langle \delta u \rangle^{\Delta} = \langle \delta u^{Left} \rangle - \langle \delta u^{Right} \rangle. \qquad (eq.\,14)$$

**Current wealth.** It should be noted that the current wealth that enters into these three models is stationary over time, fixed at the level obtained from the end of the passive phase. This is because changes to wealth are not realised until the end of the day, which means that all outcomes are hidden from the subject at the time decisions are being made. Whilst it is possible in principle to update one's expected wealth as a function of the decisions already made, this is computationally implausible, especially under the demanding cognitive constraints of the task. To compute expected wealth for a given trial, past choices have to be recalled, and integrated over all possible outcomes. This integration quickly becomes computationally implausible, especially for multiplicative condition which must take into account all of the possible wealth trajectories up to the given point in time.

**Stochastic choice function.** The stochastic choice function is identical for all models under consideration, and is comprised of a logistic function:

$$\theta\big(\langle \delta u \rangle^{\Delta}\big) = \frac{1}{1 + e^{-\beta \langle \delta u \rangle^{\Delta}}}, \qquad (eq.\,15)$$

where $\beta$ is a sensitivity parameter that determines the sensitivity of the choice probability to differences in the expected change in utility between the two gambles, and where $\theta$ evaluates to the probability of choosing the left-hand gamble. For clarity of presentation we suppress subscripts and superscripts that denote model, and subject specific parameters (Fig. 4a). Note that $\beta$ is free to vary over both subjects and condition for all three models, and thus there are two sensitivity parameters per subject, for each of the three utility models. Allowing the sensitivity parameter to change with the dynamic, allows any potential scaling differences in the change of wealth, to be accommodated in the stochasticity of the choices.

**Sampling procedures.** The Bayesian modelling affords computation of full probability distributions of parameters, rather than only point estimates which ignore the uncertainty with which parameters are estimated. Via its hierarchical structure, individuals are modelled as coming from group-level distributions, such that information from the group informs the estimation of the individual, and constrains extreme values that might be estimated with uncertainty[39]. To this end Monte-Carlo Markov Chain sampling was performed via JAGS(v4.03), called from MATLAB™ (v9.4.0.813654 R2018a, Mathworks®, mathworks.com) via the interface MATJAGS (v1.3, psiexp.ss.uci.edu/research/programs_data/jags). For all models we used: a burn-in > 500, $10^4$ samples per chain, and 10 chains for (Model recovery & parameter estimation) and 4 chains for (model selection and parameter recovery). Convergence was established via monitoring R-hat values 1 to 1.01. The sampling procedures were efficient, as indicated by low autocorrelations of the sample chains, R-hat values, and visual inspections of the chain plots.

**Model selection.** The three utility models were estimated via a single hierarchical latent mixture (HLM) model. Whilst these utility models are submodels of the HLM, for consistency we call them utility models. The HLM model is depicted graphically in Fig. 4a and with distributional and structural equations detailed listed in 4b. The sensitivity parameter $\beta$ parameter is common to all three utility models and is free to vary by subject and by condition, to accommodate any differences in the scaling of wealth changes. Following Nilsson and colleagues[39] we set weakly informative hyperpriors, such that the group mean of $\beta$ was certain to lie in an interval that ranges from 0.1 to ~30. Assuming an uninformative uniform hyperprior distribution for the lognormal group means, this translates to hyperpriors distributed as: $\mu_c^\beta \sim$ Uniform(-2.3, 3.4). We assigned uninformative uniform hyperpriors for the lognormal standard deviations $\sigma_c^\beta \sim$ Uniform(0.01, 1.6) where 1.6 is the approximate standard deviation of a uniform distribution that ranges from −2.30 to 3.4. *Time optimal utility model:* Specified as a restricted isoelastic model, with a population mean risk aversion $\mu^\eta$ fixed to 0 for additive and 1 for multiplicative dynamics. Assuming uninformative uniform hyperpriors $\sigma_c^\eta \sim$ Uniform(0.01, 1.6) for the standard deviations of the normally distributed risk aversion parameters means that only the dispersion around the [0,1] coordinate in risk aversion space is free to vary (Fig. 3c). *Prospect theory utility model:* has three further free parameters. For risk preferences it has one $\alpha$ parameter each for gains and losses, both are constrained to be lie between 0 and 1, here assumed to each come from a lognormal distribution, with an uninformative uniform hyperprior distribution on the lognormal group means and standard deviations $\mu^\alpha \sim$ Uniform(-2.3, 0) and $\sigma^\alpha \sim$ Uniform(0,1.6). The third parameter is the loss aversion parameter λ, which we assumed to lie on an interval from 1 and 5, and thus we set equivalent non-informative uniform hyperpriors on the lognormal group means and standard deviations $\mu^\lambda \sim$ Uniform(0, 1.6) and $\sigma^\lambda \sim$ Uniform(0, 1.6). *Isoelastic utility model:* Assuming uninformative uniform hyperpriors for the population mean of the risk aversion parameter $\mu^\eta \sim$Uniform(-2.5, 2.5) and $\sigma^\eta \sim$ Uniform(0,1.6) for the standard deviations of the normally distributed risk aversion parameters. *Latent mixtures of utility models:* Finally, the modelling of latent mixtures of models via indicator variables, allows model comparison between qualitatively different, as well as nested utility models, within one superordinate model[16]. The model indicator variable $z$ was set with non-informative uniform priors and was free to vary by subject. This represents our agnosticism toward which utility model is best under variable dynamics. The posterior model probabilities (Fig. 4c), estimated model frequencies (Fig. 4e) and the protected exceedance probabilities (Fig. 4f) were estimated via the Variational Bayesian Analysis toolbox[40] ([mbb-team.github.io/VBA-toolbox/](mbb-team.github.io/VBA-toolbox/)).

**Parameter estimation.** Via the hierarchical model depicted in Fig. 3a, we estimated the posterior distribution of risk aversion parameters for a single dynamic-specific isoelastic utility model, given the choice data. This model is an isoelastic model in which the risk aversion parameter is free to vary over dynamics, as well as over subjects. It is specified to be the same as the isoelastic utility model used in the model selection, except here the risk aversion parameter is estimated condition-wise, and there are no other utility models or latent model indicator variables.

**Acknowledgements.** We thank Ole Peters, Alex Adamou, Yonatan Berman, Mark Kirstein, Tobias Andersen, Jason Collins, Peter Dayan, Brad Cameron, Chris Merrill, Adam Goldstein, Alex Imas, Ilari Lehti, and Sven Resnjanskij for helpful discussions. Thank you to Félix Hubert for making Supp. Fig. 6. Funding: HS (Lundbeck Foundation Grant of Excellence "ContAct" ref: R59 A5399 ; Novo Nordisk Foundation Interdisciplinary Synergy Programme Grant "BASICS" ref: NNF14OC0011413; 5-year professorship in precision medicine sponsored by Lundbeck Foundation ref: R186-2015-2138), O.J.H (Lundbeck Foundation, ref: R140-2013-13057; Danish Research Council ref: 12-126925), DM (Novo Nordisk Foundation project grant, ref: NNF16OC0023090).

**Contributions.** OH & DM conceived of the study, OH, FR, HS & DM designed the study, FR collected the data, OH & DM & MK performed modelling, OH performed analyses and wrote the paper. All authors contributed to the interpretation, edited, and approved the final version.

**Conflict of Interest.** HS has received honoraria as speaker from Sanofi Genzyme, Denmark and Novartis, Denmark, as consultant from Sanofi Genzyme, Denmark and as senior editor (NeuroImage) from Elsevier Publishers, Amsterdam, The Netherlands. HRS has received royalties as book editor from Springer Publishers, Stuttgart, Germany.

**a**  Multiplicative growth rates for outcomes on Day$^\times$

| Fractal# | #1 | #2 | #3 | #4 | #5 | #6 | #7 | #8 | #9 |
|---|---|---|---|---|---|---|---|---|---|
| Growth factor per trial | 0.447 | 0.546 | 0.668 | 0.818 | 1.000 | 1.223 | 1.496 | 1.830 | 2.239 |
| Exponential growth per trial | -0.806 | -0.604 | -0.403 | -0.202 | 0.000 | 0.202 | 0.403 | 0.604 | 0.806 |

Additive growth rates for outcomes on Day$^+$

| Fractal# | #10 | #11 | #12 | #13 | #14 | #15 | #16 | #17 | #18 |
|---|---|---|---|---|---|---|---|---|---|
| Growth increment per trial | -428 kr | -321 kr | -214 kr | -107 kr | 0 kr | 107 kr | 214 kr | 321 kr | 428 kr |

**b**  Time average multiplicative growth rates for gambles on Day$^\times$

| Fractal# | #1 | #2 | #3 | #4 | #5 | #6 | #7 | #8 | #9 |
|---|---|---|---|---|---|---|---|---|---|
| #1 |  | -0.705 | -0.604 | -0.504 | -0.403 | -0.302 | -0.202 | -0.101 | 0.000 |
| #2 |  |  | -0.504 | -0.403 | -0.302 | -0.202 | -0.101 | 0.000 | 0.101 |
| #3 |  |  |  | -0.302 | -0.202 | -0.101 | 0.000 | 0.101 | 0.201 |
| #4 |  |  |  |  | -0.101 | 0.000 | 0.101 | 0.202 | 0.302 |
| #5 |  |  |  |  |  | 0.101 | 0.202 | 0.302 | 0.403 |
| #6 |  |  |  |  |  |  | 0.302 | 0.403 | 0.504 |
| #7 |  |  |  |  |  |  |  | 0.504 | 0.604 |
| #8 |  |  |  |  |  |  |  |  | 0.705 |
| #9 |  |  |  |  |  |  |  |  |  |

Time average additive growth rates for gambles on Day$^+$

| Fractal# | #10 | #11 | #12 | #13 | #14 | #15 | #16 | #17 | #18 |
|---|---|---|---|---|---|---|---|---|---|
| #10 |  | -374.5 | -321.0 | -267.5 | -214.0 | -160.5 | -107.0 | -53.5 | 0.0 |
| #11 |  |  | -267.5 | -214.0 | -160.5 | -107.0 | -53.5 | 0.0 | 53.5 |
| #12 |  |  |  | -160.5 | -107.0 | -53.5 | 0.0 | 53.5 | 107.0 |
| #13 |  |  |  |  | -53.5 | 0.0 | 53.5 | 107.0 | 160.5 |
| #14 |  |  |  |  |  | 53.5 | 107.0 | 160.5 | 214.0 |
| #15 |  |  |  |  |  |  | 160.5 | 214.0 | 267.5 |
| #16 |  |  |  |  |  |  |  | 267.5 | 321.0 |
| #17 |  |  |  |  |  |  |  |  | 374.5 |
| #18 |  |  |  |  |  |  |  |  |  |

**Supplementary Figure 1 | Growth rates for outcomes and gambles. a**, on Day$^\times$ (upper table) a growth factor is the factor by which current wealth is multiplied when a given stimulus is encountered in the passive session. The effect of each stimulus can thus be expressed as a multiplicative growth rate (in units of *growth factor per trial*). Computing the natural logarithm of the growth factor per trial gives a continuous growth rate (in units of *% change per trial*). On Day$^+$ (lower table), growth increments are the additive amounts by which wealth changes, and thus the growth rate is an additive growth rate (in units of *DKK per trial*). **b,** each gamble is comprised of two different possible outcomes, here denoted in terms of pairs of stimuli. Each cell shows the time average growth rate associated with each gamble in the space of possible gambles. The cells with red text indicate the 16 gambles that were presented in the active sessions. For Day$^\times$ (upper) the time average multiplicative growth rates of each gamble have units *% change per trial*. For Day$^+$ (lower) the time average additive growth rates have the units of *DKK per trial*.

**Choice proportion analysis Day$^+$.** In the following H0 denotes the null hypothesis, H- to denotes the alternate hypothesis specifying values less than a reference value, and H+ to denote the equivalent for values above a reference value. Bayes factors obeys the same notation: BF$_{-0}$ denotes a Bayes factor for H- over H0, BF$_{0-}$ for H0 over H-, and so on. To assess choice proportions on Day$^+$ we performed a one-sample Bayesian t-test in which we assign effect sizes a zero-centred Cauchy prior with scale 0.707 ($\frac{1}{\sqrt{2}}$). The fat-tailed Cauchy distribution is used because it fulfils particular criteria[41,42]. Of interest is the posterior distribution for the underlying choice proportion $CP_{log}$. The resulting posterior distribution, which is concentrated near 0.5, with a central 95% credible interval of 0.395 to 0.591. The alternative hypothesis (H-) is relatively informative insofar as it states that $CP_{log}$ is lower than 0.5, but that values of $CP_{log}$ close to 0.5 are more likely than those values far below it (H- : $0 < CP_{log} < 0.5$) as seen in Fig. 2b which shows the one-sided prior and posterior distribution for the effect size of $CP_{log}$ under the informative H-. Correspondingly, the null hypothesis (H0) states that agents will choose with respect to the linear utility less often than in favour of log utility, and thus predicts that the choice proportion in favour of log utility will be larger than 0.5 (H0: $CP_{log} > 0.5$). A one sample Bayesian t-test revealed a $BF_{0-}$ of 3.678, which indicates the null hypothesis is nearly 4 times more likely than the alternative, which can be classed as moderate evidence. As shown in Fig. 2b, compared to the prior distribution, the posterior distribution is more concentrated near an effect size of 0. For robustness checks, the effect of different prior widths (wide and ultrawide priors, scale factors 1 and $\sqrt{2}$, respectively) can be seen in Fig. 2c&d, which show that they do not effectively change this interpretation. In conclusion, this indicates moderate evidence that under additive dynamics, choices in favour of linear utility were not more likely than those in favour of log utility.

**Choice proportion analysis Day$^\times$.** As above, to assess choice proportions on Day$^\times$ we performed a one-sample Bayesian t-test in which we assign effect sizes a zero-centred Cauchy prior, with scale 0.707. Of interest is the posterior distribution for the underlying choice proportion $CP_{log}$. The resulting posterior distribution is concentrated near 0.7, with a central 95% credible interval for $CP_{log}$ that ranges from 0.625 to 0.812. The alternative hypothesis is relatively informative and states that $CP_{log}$ is higher than 0.5, but that values of $CP_{log}$ close to 0.5 are more likely than values far above it (H+ : $1 > q > 0.5$) as seen in Fig. 2e which shows the one-sided prior and posterior distribution for the effect size of $CP_{log}$ under the informative H+. The null hypothesis states that agents will not choose with respect to the log utility more often, and thus predicts that the choice proportion in favour of the log utility will be smaller than 0.5 (H0: $CP_{log} < 0.5$). A one sample Bayesian t-test revealed a Bayes Factor for the data being ~460 times more likely under H+ than under H0, which is classed as extreme evidence. As shown in Fig. 2e, compared to the prior distribution, the posterior distribution is concentrated near an effect size of 1. Robustness checks and sequential analysis can be seen in Fig. 2f-g, and do not effectively change this interpretation. In conclusion, this indicates extreme evidence that under multiplicative dynamics, choices in favour of log utility are more likely than those in favour of linear utility.

**Effect of dynamic on choice proportion.** To assess within subject changes in choice proportion following the different dynamics, we performed a Bayesian paired t-test in which we assign effect sizes a zero-centred Cauchy prior with scale 0.707. Of interest is the posterior distribution for the between-dynamic difference in choice proportion $\Delta CP_{log}$. The resulting posterior distribution is concentrated near a proportion difference of 0.23, with a central 95% credible interval for $CP_{log}$ that ranges from 0.099 to 0.351. The null hypothesis states that agents will not change their choice proportion under different dynamical conditions, and thus predicts that the choice proportion will be equal for each condition (H0 $\Delta CP_{log} = 0$). The alternative hypothesis is relatively informative and states that $\Delta CP_{log}$ is larger than 0, but that values of $\Delta CP_{log}$ close to 0 are more likely than values far above it (H+ : 1> $\Delta CP_{log}$ > 0) as seen in Fig. 2i which shows the one-sided prior and posterior distribution for the effect size of $\Delta CP_{log}$ under H+. The paired Bayesian t-test revealed a Bayes factor of 52.376, which indicates the alternate hypothesis is around 50 times more likely than the null, which can be classed as very strong evidence. As shown in Fig. 2i, compared to the prior distribution, the posterior distribution is concentrated near an effect size of 0.8. Robustness checks and sequential analyses can be seen in Fig. 2j&k, and do not effectively change this interpretation. In conclusion, we find that gamble dynamics have a very strong effect on choice frequencies, with the gamble dynamics moving choice frequencies in the direction predicted by time optimality.

**a**

Model Comparison

| Models | P(M) | P(M|data) | $BF_M$ | $BF_{01}$ | error % |
|---|---|---|---|---|---|
| Dynamic | 0.050 | 0.222 | 5.427 | 1.000 | |
| Dynamic + Age | 0.050 | 0.162 | 3.660 | 1.376 | 1.857 |
| Dynamic + Gender (0=f) | 0.050 | 0.091 | 1.895 | 2.450 | 2.346 |
| Dynamic + *First | 0.050 | 0.091 | 1.893 | 2.452 | 1.807 |
| Dynamic + Age + Gender (0=f) | 0.050 | 0.085 | 1.772 | 2.604 | 2.618 |
| Dynamic + *First + Dynamic ∗ *First | 0.050 | 0.076 | 1.569 | 2.912 | 2.006 |
| Dynamic + *First + Age | 0.050 | 0.065 | 1.318 | 3.424 | 2.024 |
| Dynamic + *First + Dynamic ∗ *First + Age | 0.050 | 0.058 | 1.166 | 3.841 | 3.221 |
| Dynamic + *First + Gender (0=f) | 0.050 | 0.038 | 0.748 | 5.862 | 2.579 |
| Dynamic + *First + Age + Gender (0=f) | 0.050 | 0.036 | 0.710 | 6.169 | 2.873 |
| Dynamic + *First + Dynamic ∗ *First + Gender (0=f) | 0.050 | 0.035 | 0.693 | 6.318 | 2.680 |
| Dynamic + *First + Dynamic ∗ *First + Age + Gender (0=f) | 0.050 | 0.033 | 0.655 | 6.662 | 3.052 |
| Null model (incl. subject) | 0.050 | 0.003 | 0.048 | 87.729 | 1.224 |
| Age | 0.050 | 0.002 | 0.030 | 139.364 | 1.847 |
| Gender (0=f) | 0.050 | 9.948e−4 | 0.019 | 223.324 | 2.213 |
| *First | 0.050 | 9.797e−4 | 0.019 | 226.765 | 1.463 |
| Age + Gender (0=f) | 0.050 | 8.032e−4 | 0.015 | 276.584 | 2.926 |
| *First + Age | 0.050 | 6.272e−4 | 0.012 | 354.236 | 2.516 |
| *First + Gender (0=f) | 0.050 | 3.950e−4 | 0.008 | 562.481 | 2.939 |
| *First + Age + Gender (0=f) | 0.050 | 3.220e−4 | 0.006 | 689.991 | 3.340 |

Note. All models include subject.

**b**

Analysis of Effects ▼

| Effects | P(incl) | P(incl|data) | $BF_{Inclusion}$ |
|---|---|---|---|
| Dynamic | 0.600 | 0.992 | 80.157 |
| *First | 0.600 | 0.434 | 0.512 |
| Age | 0.500 | 0.442 | 0.793 |
| Gender (0=f) | 0.500 | 0.321 | 0.473 |
| Dynamic ∗ *First | 0.200 | 0.203 | 1.017 |

**Supplementary Figure 2 | Model comparisons and analysis of effects for choice proportions. a,** the table of model probabilities, Bayes factors and error terms, for a repeated measures ANOVA on the choice proportions for discrepant trials. The meanings of each column are described in the text. **b,** the inclusion probabilities for all factors of interest across all models, along with the Bayes factors for their inclusion.

**Repeated measures ANOVA for choice proportions.** We conducted a Bayesian repeated measures ANOVA on the choice frequency data, with gender, age, and order of testing (*First) as between subject factors, and dynamic as within subject factors. We used the default prior options for the effects (r = 0.5 for the fixed effects, prior scale factor 0.707). To assess the robustness of the result, we also repeat the analysis over wide and ultrawide priors. The 'Model Comparison' table (Supp. Fig. 2a) gives the results with respect to the different models that are compared. The models that are considered are all possible models including interactions of factors. The table lists all of the models, and the corresponding Bayes factors, where the best performing model (here, this is the model that includes only the dynamic factor) is compared to all the other models. The column of $BF_{01}$ shows that the data are ~6.6 times more likely under the model with only the dynamic, than under the full model (i.e., the model with age, gender and order of testing, and their interactions). The column P(M) lists the prior model probabilities, which are held uniform across all the models. The column P(M|data) lists the posterior model probabilities. The column $BF_M$ lists the comparisons between the best model (dynamic factor only) and each other model. The 'Analysis of Effects' (Supp. Fig. 2b) gives Bayes factors for the inclusion of each effect that appears in at least one model. For each effect, the BF$_{inclusion}$ column reflects how well the effect predicts the data by comparing the performance of all models that include the effect to the performance of all the models that do not include the effect. For the gamble dynamic, there is very strong evidence in favor of its inclusion (BF$_{inclusion}$ > 80), whereas for all other factors there is either evidence against their inclusion, or only anecdotal evidence for their inclusion. In conclusion, compared to other factors and covariates, gamble dynamics have a uniquely strong effect on choice frequencies.

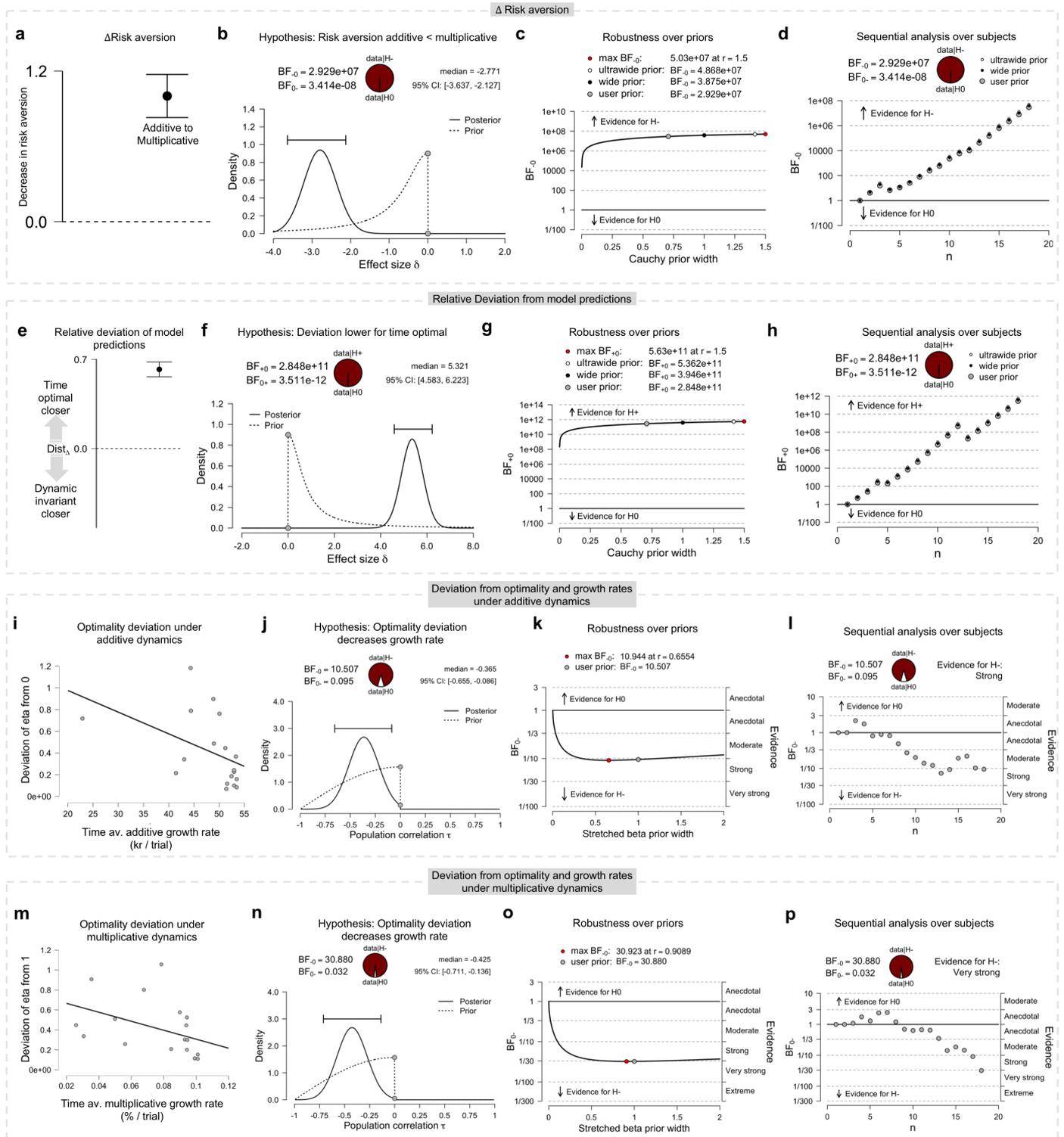

**Supplementary Figure 3 | Descriptive statistics, priors & posteriors of hypothesis test, robustness tests and sequential analyses. a-d,** effect of dynamics on changing risk aversion parameters. **h-k,** comparison of the deviation of each model predictions of risk aversion parameters to those observed. **l-o,** effect of deviating from

time optimality on the time average growth rates of subjects' choices, under additive dynamics. **p-s,** equivalent effect under multiplicative dynamics.

**Gamble dynamics exert strong effects on risk aversion parameters.** To assess within subject changes in $\eta$ following the different dynamics, we performed a Bayesian paired t-test in which we assign effect sizes a zero-centred Cauchy prior with scale 0.707. Of interest is the posterior distribution for the between-dynamic difference in $\eta$, denoted $\Delta \eta$. When comparing the $\eta$ of the multiplicative to the additive, the resulting posterior distribution is concentrated near a decrease of 1.01, with a central 95% credible interval for $\Delta \eta$ that ranges from 0.829 to 1.172 (Supp. Fig. 3a). The null hypothesis states that agents will not change their risk aversion under different dynamical conditions, and thus predicts that $\eta$ will be equal for each condition (H0: $\Delta \eta$ = 0). The alternative hypothesis is relatively informative and states that $\Delta \eta$ is less than 0, but that values close to 0 are more likely than values far below it (H- $\Delta \eta$ < 0) as seen in Supp. Fig. 3b which shows the one-sided prior and posterior distribution for the effect size of $\Delta \eta$ under H-. The paired Bayesian t-test revealed a Bayes factor of $2.9 \times 10^7$, which indicates extreme evidence in favour of the alternate hypothesis. As shown in Supp. Fig. 3b, compared to the prior distribution, the posterior distribution is concentrated near an effect size of -3. Robustness checks over different prior widths can be seen in Supp. Fig. 3c&d, and do not effectively change this interpretation. Descriptive statistics for the $\eta$ parameter are in Supp. Fig. 4a. In conclusion, there are strong effects of gamble dynamics on risk aversion, with the multiplicative dynamics increasing estimated risk aversions, compared to additive dynamics.

**Estimates of risk aversion are closer to predictions of time optimal model.** To establish whether the $\eta$ values are closer in "$\eta$-space" to the predictions of the time optimal or dynamic-invariant models, we computed the Euclidean distances of each subjects MAP estimate to each of the models predicted coordinate(s): In the time optimal case this is simply the distance to the [0,1] coordinate, whereas for the dynamic invariant utility model this is the shortest distance to the main diagonal (Fig. 3c). We are interested to test whether these distances to the model predictions are smaller under the time optimal model, and thus we compute the difference in distance as $Dist_\Delta = Dist_{invariant} - Dist_{time}$. $Dist_\Delta$ had a mean of 0.65 (Supp. Fig. 3e) indicating the time optimal model was closer in its predictions. To test this, we performed a Bayesian paired t-test in which we assign effect sizes a zero-centred half-Cauchy prior with scale 0.707. Of interest is the posterior distribution for the effect sizes of $Dist_\Delta$. The resulting posterior distribution is concentrated near a median effect size of 5.321, with a central 95% credible interval that ranges from 4.583 to 6.223 (Supp. Fig. 3f). The alternative hypothesis is relatively informative and states that difference in distances will be positive, but that values close to 0 are more likely than values far above it (H+: $Dist_\Delta$ > 0) as seen in Supp. Fig. 3f which shows the one-sided prior and posterior distribution for the effect size of $Dist_\Delta$ under H+. The null hypothesis states that the distance of the data to the predictions is larger for the dynamic invariant model than for the time optimal model, and thus predicts that the difference in distances will be negative (H0 : $Dist_\Delta$ < 0). The paired Bayesian t-test revealed a Bayes factor of $2.8 \times 10^{11}$, which indicates extreme evidence in favour of the alternate hypothesis. As shown in Supp. Fig. 3f, compared to the prior distribution, the posterior distribution is concentrated near an effect size of 5.3. Robustness checks and sequential analyses over different prior widths can be seen in Supp. Fig. 3g&h, and do not effectively change this interpretation. In conclusion, there is extreme evidence that the estimated risk aversions are closer to the prediction of the time optimal model than a model which assumes no dynamic specific changes in risk aversion.

**Deviations from time optimality correlates negatively with time average growth rates.** We conducted a Bayesian correlation analysis for the relation between the deviation of the estimated risk aversion from time optimality, and the time average growth rate achieved by the participants' choices. We used a default prior (as specified in JASP software) which yields a uniform distribution on Kendall's $\tau$ [43]. We focus on hypothesis testing, specifying a one-sided alternative hypothesis which posits a negative correlation between deviation from time optimality and time average growth rate, compared to the null hypothesis that postulates that the correlation is non-negative. The Bayes factor for each correlation quantifies the evidence in favor of a negative correlation. Negative correlations were found for both additive dynamics ($BF_{-0}$ = 30.88, $BCI_{95\%}$ [-0.656, -0.068]) and multiplicative dynamics ($BF_{-0}$ = 10.51, $BCI_{95\%}$ [-0.711, -0.131]). These correlations yielded $BF_{-0} > 10$, indicating strong evidence in favor of the alternative hypothesis that postulates the presence of a negative correlation. The posterior distributions for each of the correlations are in Supp. Fig. 3j&n, and scatterplots between the variables are in Supp. Fig. 3i&m, including a fitted linear regression line. Robustness checks and sequential analyses over different prior widths are shown in Supp. Fig. 3k-l and o-p. A table of statistics, including $\tau$ estimates is in Supp. Fig. 4d.

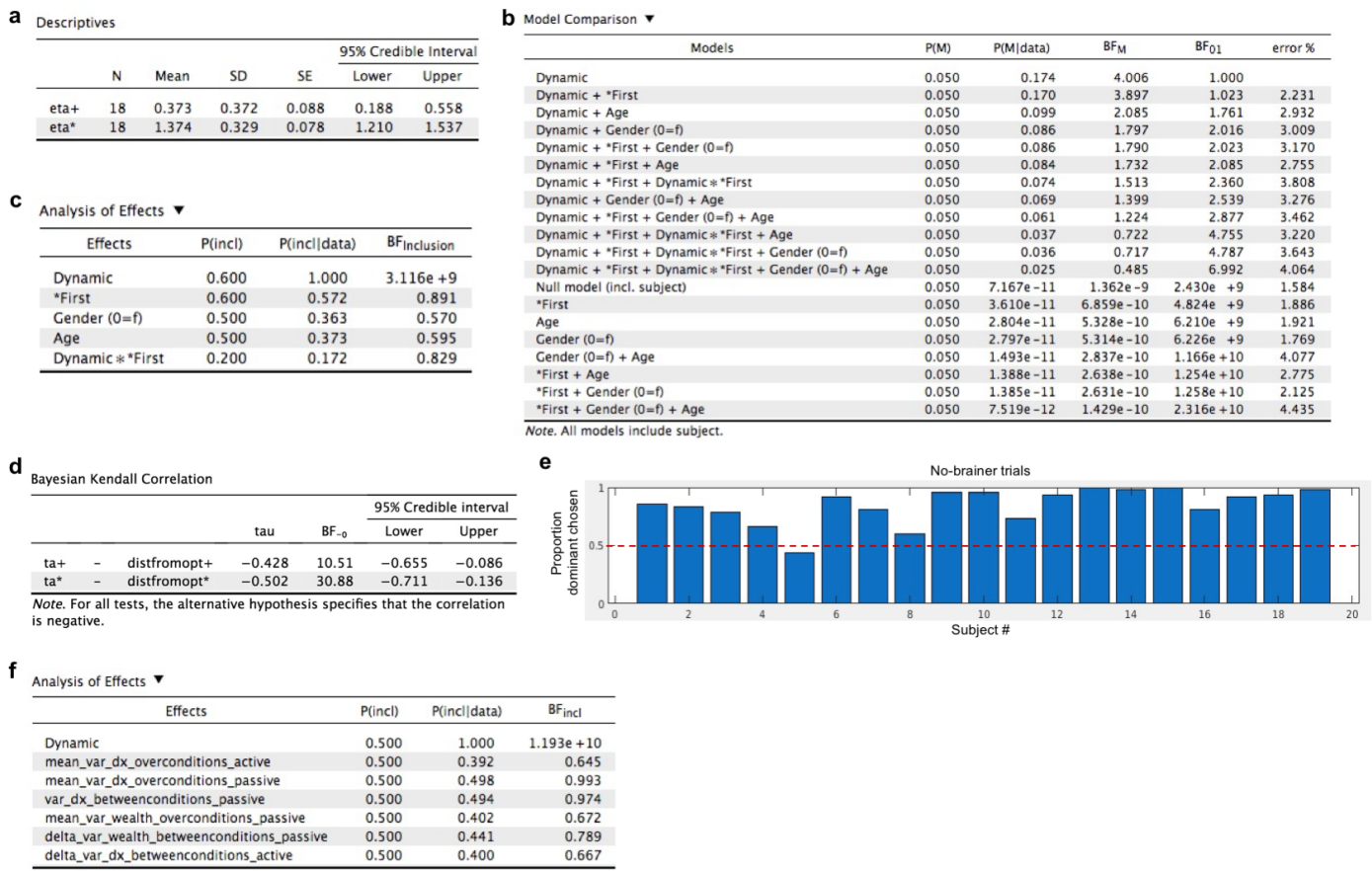

**Supplementary Figure 4 | Tables and no-brainers. a,** descriptive statistics for the risk aversion parameter, $\eta$. SD - standard deviation; SE - standard error mean. **b,** table of models compared in repeated measures ANOVA for risk aversions. Column headings described in main text. **c,** table of Analysis of effects shows the Bayes factors for the

inclusion of each factor across all of the models in b. P(incl) indicates the prior probability of each effect across all models. $BF_{Inclusion}$ is the Bayes factor for the inclusion of that factor, comparing all models with vs. without that factor. **d,** table for Kendall correlation, where correlations between time averages (ta) and distances from optimal risk aversion parameters ('distfromopt') are tabulated. **e,** proportion of correct responses (dominant chosen) in the no-brainer trials. Red line indicates chance performance. **f,** Analysis of effects shows Bayes factors for the inclusion of each factor, including both the dynamic, and the covariates derived from the variances in wealth and changes in wealth.

**Repeated measures ANOVA shows strong effect of gamble dynamics on risk aversion.** We conducted a Bayesian repeated measures ANOVA on the risk aversion parameter $\eta$, with gender, age, and order of testing as between subject factors, and dynamic as a within subject factor. We used the default prior options for the effects (r = 0.5 for the fixed effects). To assess the robustness of the result, we also repeat the analysis for different widths of prior. The 'Model Comparison' table in Supp. Fig. 4b gives the results with respect to the different models. The models that are compared are all possible combinations of factors including interactions of factors. The table lists all of the models, and the corresponding Bayes factors, where the best performing model (here, the model that includes only the dynamic factor) is compared to all the other models. The column of $BF_{01}$ shows that the data are ~7 times more likely under the model with only the dynamic, than under the full model (i.e., the model with age, gender and order of testing, and their interactions). The column P(M) lists the prior model probabilities, which are held uniform across all the models. The column P(M|data) lists the posterior model probabilities. The column $BF_M$ indicates how many times the best model is compared to each other model. The 'Analysis of Effects' (Supp. Fig. 4c) gives Bayes factors for the inclusion of each factor that appears in at least one of these models. For each factor, the $BF_{Inclusion}$ column reflects how well the effect predicts the data by comparing the performance of all models that include the factor to the performance of all the models that do not include the factor. For the factor representing the gamble dynamic (Dynamic), there is extreme evidence in favor of its inclusion ($BF_{Inclusion} > 100$), whereas for all other factors there is evidence against their inclusion. In conclusion, compared to other factors, the gamble dynamic has a uniquely strong effect on risk aversions.

**Idiosyncratic wealth trajectories exert no systematic effect on risk aversions.** Since the passive phase was designed to have stochastic paths, subjects can end the passive phase with different wealths, and also have experienced different volatilities (variance of wealth changes). We tested whether these differences between both subjects and conditions, could account for the differences in risk aversions observed. To this end, we performed a rmANOVA with eta as dependent variable, and dynamic as repeated measures factor. The following subject-wise covariates were added for changes in wealth: mean variance across both conditions for passive phase ('mean_var_dx_overconditions_passive'); difference in variance between conditions for passive phase ('delta_var_dx_betweenconditions_passive'); mean variance across both conditions for active phase ('mean_var_dx_overconditions_passive'); difference in variance between conditions for active phase ('delta_var_dx_betweenconditions_passive'). Two further subject-wise covariates were added for in-game wealth itself: mean variance across conditions for passive phase ('mean_var_wealth_overconditions_passive'); difference in variance between conditions for passive phase ('delta_var_wealth_betweenconditions_passive'). Note that Supp. Fig. 4e shows that the Bayes Factor for the inclusion of the dynamic factor is extreme, whereas all other covariates are below 1. This indicates that there is not even moderate evidence that adding wealth covariates to the model improves its predictive adequacy.

**Parameter recovery.** To evaluate whether the model estimation methods were capable of recovering approximate parameter estimates, we performed a parameter recovery simulation in which we subjected our estimation procedures to synthetic data for which ground truth parameter values were set a priori. Supp. Fig. 5a shows the correspondence between the estimates of risk aversion parameters and the ground truth values used in simulating synthetic agents. Agents were simulated to have all pairwise combinations of $\eta$ values of [-0.5,0,0.5,1,1.5] for additive and multiplicative dynamics. 20 subjects were simulated for each parameter combination, and then the same parameter estimation procedures were applied to visualise the recovery of parameters as used in Fig. 3. This includes the estimation of both $\beta$ and $\eta$ parameters. Supp. Fig. 5a shows a subset of this space most relevant to the key results of this paper. The fact that $\eta$ can be recovered accurately shows that it cannot be adequately captured via other parameters such as the sensitivity parameter $\beta$. This is also evident from the relative precision of the posterior $\eta$ values observed in the original data in Fig. 3.

**Model recovery.** To evaluate whether the model selection methods were capable of recovering the set of utility models tested, we performed a model recovery simulation in which we subjected our estimation procedures to choices made by synthetic agents for which ground truth model values were set a priori. Supp. Fig. 5b shows the correspondence between the posterior inclusion probabilities for each utility model and the ground truth identities of the utility models used in simulating synthetic agents. The first seven subjects were synthesised as prospect theory agents (with same parameters for additive and dynamic sessions), the next seven subjects as isoelastic utility agents (again with same parameters for both sessions), and finally the last seven subjects were time optimal agents.

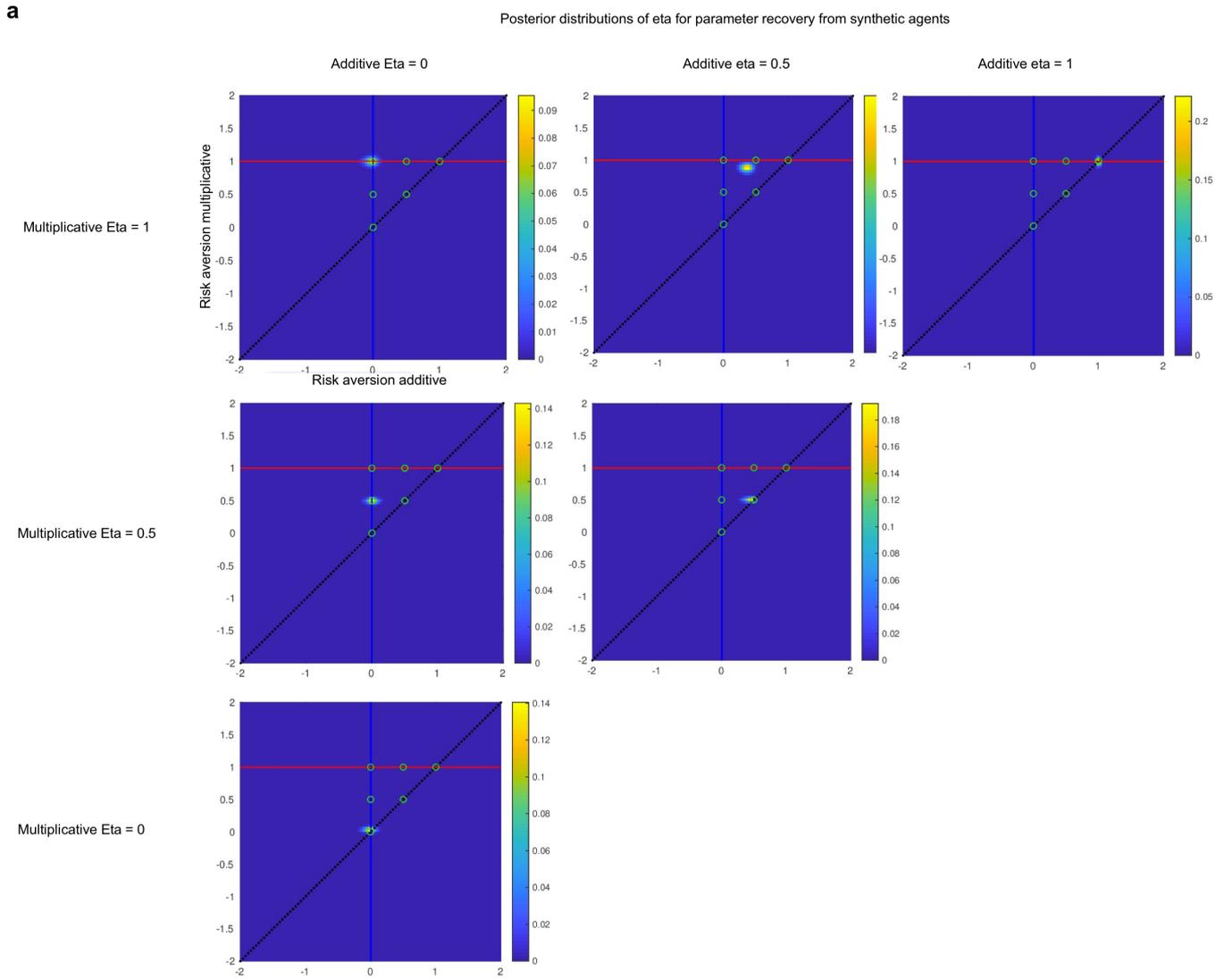

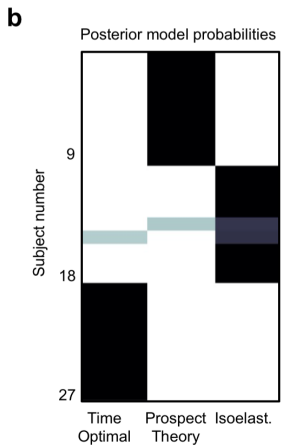

**Supplementary Figure 5 | Parameter and Model recovery. a,** parameter recovery for several populations of synthetic agents with different combinations of $\eta$ parameters. Each panel shows the posterior $\eta$ distribution, marginalised over subjects, estimated via the same model and code as the real data shown in Fig. 3. **b,** model recovery for three different groups of synthetic agents, for the three models compared. Color range shows posterior model probabilities as in Fig. 4c (black = 1, white = 0). Posterior model probabilities map strongly onto the ground truth model identities, insofar as the first nine agents were synthesised via a time optimal model, the next nine from a prospect theory model, and the final nine from an isoelastic utility model. Note that the isoelastic

agent obtains small posterior model probabilities in the range of ~0.05 for the two other models, for only 2/9 parameter values.

**Synthetic agents.** Fig. 1e-f shows wealth trajectories of synthetic agents repeatedly playing the set of 144 different gambles used in this paradigm. No-brainers were not included. 12 different agents were synthesised, comprising the three model classes: 9 variants of prospect theory agents, comprising all possible combinations of $\lambda$ {1, 2, 3} and $\alpha$ {0.3, 0.6, 0.9} (identical for both gains and losses), and 2 variants of isoelastic agents $\eta$ {0, 1}. Each prospect theory and isoelastic agent had the same parameters for both additive and multiplicative dynamics. The time optimal agent is a special case, having linear utility under additive dynamics, and logarithmic utility under multiplicative dynamics. The trajectories were computed over several timescales, hour, day, week, year. At a long enough timescale noise is removed by the passage of time, and the time average growth rates of the different agents become apparent. In this artificial environment, agents were playing trials every 9.5s, continuously. Over the duration of a week, a reasonable approximation to the time average growth rate is typically revealed (Fig. 1e-f).

## Supplementary Discussion.

**House money.** Another validity issue for economic experiments is the issue of "house money", which pertains to whether subjects behave differently if they perceive the money not to be entirely theirs[22]. These effects are typically circumvented by the pre-endowment of money, or by having subjects work in order to feel ownership of the money. Both of these strategies were implemented here, insofar as subjects knew days ahead of their upcoming endowment, and they were pre-endowed approximately 90 minutes prior to decision-making. The sense of ownership was enhanced by subjects actively causing changes in this wealth via 338 button presses and observing the resulting fluctuations for a total of ~60 minutes. Further, house money as a putative effect does not explain the condition-specific changes in risk aversion observed, nor their bivariate alignment with the time optimal strategy (Fig. 3e).

**Cognitive considerations.** It is notable that subjects were able to perform the task under challenging cognitive conditions. Gambles were chosen based on the participant's memory of the stimuli from the previous passive session, up to 60 min ago, with choices being made every ~10s for ~1 hour of testing per day in a noisy environment. Nearly all participants (18 of 19) tested could choose dominant gambles above chance in the No-brainer trials (Supp. Fig. 4e). The degree to which subjects could approximate the time optimal strategy indicates that they had a relatively high-fidelity magnitude representation of the underlying growth increments and factors. Though time optimality was a reasonable approximation of the observed data, the estimated risk aversion parameters were systematically biased to be more risk averse than predicted by time optimality (Fig. 3p). This may be due to noise in the sensory or mnemonic encoding of the stimuli. Such sources of noise may increase the apparent risk aversion due to the uncertainty it adds to each gamble.

**Dynamical utility models.** The models tested so far were all static, in the sense that they do not incorporate any anticipation of future gambles or wealth trajectories. This is because the game is effectively a single period game in which the 10 randomly selected outcomes are realised at once at the end of the game, with no intermediate wealth updating. Nevertheless, one strategy that subjects could take is to plan ahead, making decisions that maximise the expected utility of the terminal wealth occurring at the end of the game. This is predicted under multi-period expected utility theory[17–19] which involves an iterative evaluation of wealth computed via dynamic programming[20]. Such models compute all terminal wealths that are possible under the different contingencies, and work backwards to derive the optimal choices, given the agent's utility function. However, such a strategy is not possible for subjects in this game. This is because the branching factor for each trial is extreme (>600 per trial, Supp. Fig. 6c), and the subject lacks critical information necessary to compute terminal wealths, such as knowing what the space of possible gamble pairs is, or even knowing the total number of trials they will face. We show that taking into account even optimistic estimates of cognitive constraints, results in search horizons that are so myopic, that the predictions of multiperiod models are scarcely different from the static versions.

**Dynamic versions of expected utility theory and prospect theory.** Multiperiod EUT models assume a single utility function for evaluating the terminal wealths, which is invariant across all settings. The subtlety is that, an experimenter estimating utility functions based on the observed choices of a multiperiod EUT agent would reveal an estimated utility function that appears to change between different dynamical settings, even though the agent is optimising the same utility function for terminal wealths. Indeed, this was proposed as a candidate model for this experiment[27] on the basis that it can be shown that risk aversion can decrease in the context of repeated additive gambles, when compared to multiplicative gambles. This will be important for evaluating expected utility models under dynamical settings, however multiperiod EUT is not a viable model of human behavior for this experimental task. Firstly, the terminal wealths are not tractable. This is because the subjects did not know the following five details necessary to perform a backward induction from terminal wealths: a) how many gambles they would face; b) what gambles they will face; c) the outcome of each gamble; d) which choices would be realised; e) what their current wealth is at the time of choosing. Secondly, even if we were to imagine that these uncertainties are not a problem for computing terminal wealths, the combinatorics of the game are computationally prohibitive, as we will show in the next two sections.

**A simplified game tree.** To illustrate the forking possibilities within the experiment, consider a much simpler experimental game, involving only two possible pairs of gambles, lasting only two trials long. Here the graph of the state space looks like Supp. Fig. 6a. From the central node, the graph branches first into two, one for each of the possible pairs of gambles (red dots), then it branches in two again for the binary choice that the agent makes (yellow dots), then it branches again to determine which of the two stimuli are realised by the fair coin (green dots). After one trial there are thus 8 possible terminal wealths (inner green circle of points), and after two trials there are 64 (outer green circle). Evaluating trees via planning is computationally expensive in terms of time, working memory, and metabolic energy. There is however some evidence humans can partially search trees of this size via a model-based cognitive system that evaluates the state based contingencies of sequences of choices[44,45]. On the available evidence, performance in tree-based search appears to fall off steeply with depth, with depths of 5 (for branching factors of 2) being the point at which chance performance is reached for discriminating best outcomes[44]. In the simplified game illustrated in Supp. Fig. 6a, subjects would be at chance performance by the time they are evaluating the terminal wealths after only two trials. According to the calculations of Goldstein[27], the change in the risk aversion parameter for a subject with logarithmic utility playing a repeated additive game, searching ahead two trials would result (for his setup) in approximately a shift of only 2% down to 0.98. This level of change in risk aversion is not sufficient to account for the changes in risk aversion that were observed (Fig. 3k). With such myopic search capacities, then the multiperiod model proposed by Goldstein[27] result in changes in risk aversion that are on the scale of those already illustrated in Fig. 3c (lower). The computational overheads grow exponentially with search depth. For instance, to search 4 trials ahead, Supp Fig. 6b shows that 1536 terminal wealths must be evaluated. Even if it were cognitively possible to evaluate this many nodes of a tree, which we assert it is not, then risk aversion would drop by only ~5%.

**The actual game tree for this experiment.** If we are to temporarily assume, what we know is impossible, that subjects know exactly the unknown gamble space and the number of trials, then the state space graph looks like that of Supp. Fig. 3c. This graph has a per trial branching factor with a lower bound of ~600 and a depth of 312. By comparison, the game Go has a branching factor of ~200 per move and an average depth of ~200. Evidence from neural decoding as subjects engage in planning tasks, suggest that the fastest state to state transitions are on the order of 40ms, which would correspond to approximately 25 evaluations per second[46]. Given that subjects have ~6 seconds between trials to engage in search, the number of graph nodes that can be evaluated per trial is well below what is necessary to plan one trial ahead. Even if subjects could summon cognitive capacities surpassing what has been previously observed, this would again result in only minor modulations to the risk aversions predicted by the static models. Finally, and perhaps most importantly, since multiperiod EUT is an extension to expected utility theory, it does not yield quantitative prediction for which utility functions the agent should be maximising at terminal wealth. On these considerations, multiperiod EUT is ill-suited to games of this scale, complexity, and epistemic uncertainty. Even if such issues are to be resolved or dismissed, they still do not provide quantitative predictions compatible with the data observed here. We also note that whilst it is possible to derive dynamical versions of prospect theory[47,48], the same fundamental limitations apply to these models. In both cases the cognitive limits on planning ahead make the predictions of the dynamic and static versions effectively the same.

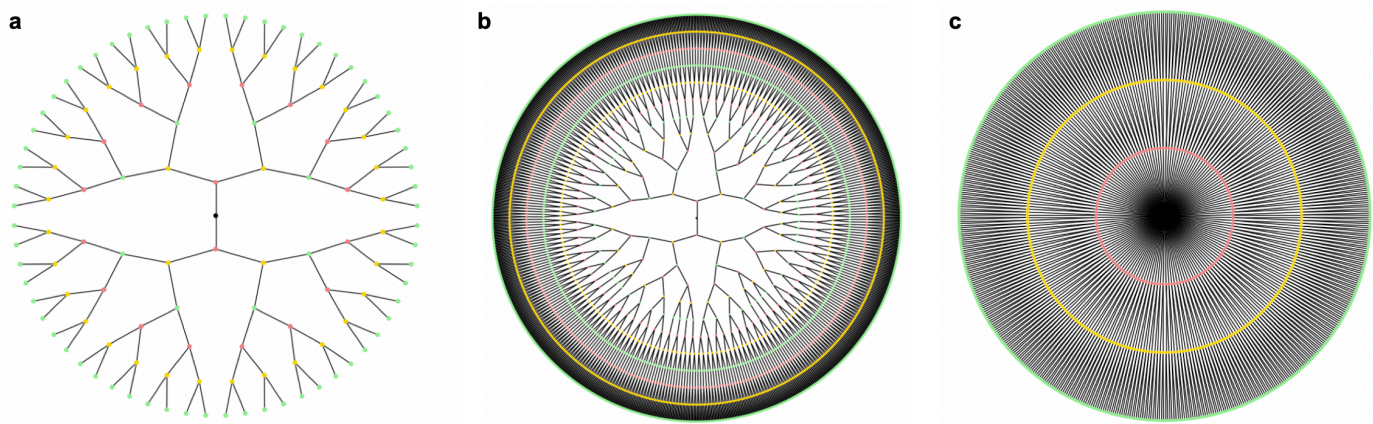

**Supplementary Figure 6 | State-space graphs for different versions of the task. a,** state-space graph of a simplified version of the game comprising only two possible gamble pairs, and with only two trials. **b,** the same state-space for four trials. **c,** The state-space for this experiment for a single trial. In all graphs a-c, red dots indicate branches between possible pairs of gambles, yellow dots indicate possible branches between choices, and green dots indicate branches between different possible outcomes.

# Supplementary Materials

## Experimental checklist

**Before subject arrives** ☐

Get paperwork (subject - on wall, MR-safety form - trays in clinic marked *kontrolskema*) ☐

Turn on the projector ☐

Locate button boxes (dual button-box, multi-color buttons, use "left") ☐

Locate tape, earplugs, & blue paper ☐

Cover stretcher paper, tear off at perforation ☐

Ensure headcoil on shelf ☐

Set up screen inside scanner ☐

Check data pie chart, if more than 3/4 delete data (only deleted data if logbook says transferred to pacs) ☐

**If first scan of the day** ☐

Arrive 30 min early ☐

Turn on scanner under quench button (~10-20min) ☐

Unlock scanner with key under quench button ☐

Turn on both computers ☐

**Greet subjects** ☐

Remember name, use frequently ☐

Fill out kontrolskema - hold on to pen ☐

Fill out subject id form ☐

Remind entering a magnet > metal attracted into scanner ☐

Remove all metal (earrings, rings, piercings, jewellery, loose change, hairties, stuff in pockets, check turnups of trousers, condom pockets, glasses, wire in bras) ☐

Remove any unnecessary objects (even if non-metal) ☐

Double check metal removal do not take their word for it, even if experienced researcher ☐

Offer hospital gown for testing (change in locker) ☐

Place removed items in locker ☐

**Training subjects** ☐

Remind subjects of game. Read them the instruction sheet ☐

Run ~20 trials outside of scanner or as many as necessary ('training.py', default 20, 5mins each for passive then active) ☐

Demo what happens if they dont press *If you don't press in time, it means the scan lasts longer* ☐

Explain time of whole experiment (3hrs total, ~60mins part 1, ~75mins part2, with breaks) ☐

**Register subject** ☐

Fill out log book, put logno as patient id ☐

On console, set patient id to the logbook no. ☐

On console, set patient surname to subject id ☐

Project name needs to be LogUtil for PACS ☐

Load LogUtil sequence from the program card ☐

**Getting ready to enter scanner room** ☐

Does S needs toilet? ☐

Explain S can talk anytime while the scanner is not running ☐

Explain intercom works like a walky-talky, while we talk, we cannot hear, wait before speaking ☐

Explain even small movements will impair data quality, this goes for moving body parts, esp. head & eye. ☐

Very very important to lie as still as possible ☐

Remember fixate numbers/cross at all times ☐

Don't form any loops with your body (no crossed legs, no touching of hands) ☐

Explain possible muscle twitches when scanner runs, all normal & not dangerous ☐

Just before entry to scanner room triple check for metal items including anyone else with you (even if senior) ☐

**In scanner room - information**

Ask them to sit on bed ☐

Inform S scanner makes different noises (bleeps, knocks, grinding, buzzing, all normal) ☐

Introduce response devices (Use "left") ☐

Attach physiological noise (pulse & resp: make sure valve points down) ☐

Ask subject to lie down on bed ☐

Show alarm bulb, loop under resp belt, explain only if urgent ☐

Earplugs ☐

Apply padding to head (slightly tight, not too tight) ☐

Head coil, ensure eyebrows align with notch ☐

Attach mirror, make sure screen roughly centre (single mirror nearest screen) ☐

Ask them if they feel comfortable / warm enough / cushion under legs? ☐

Doublecheck everything okay before entering bore ☐

Hit autoloader ☐

Explain leaving room, and will talk on other side ☐

Check intercom & <u>ask to squeeze alarm bulb</u> ☐

Turn off lights in scanner room (switch 2 underneath phys noise in control room) ☐

**Data collection pre-break**

0. Start game so it waits for trigger. <u>Do not move cursor of test computer while game is playing</u> ☐

*1. We now run a localiser scan to localise your head, It will only be 30s, expect buzzing-grinding-knocking-bleeping noises, nothing to do just relax* ☐

2. AAhead_scout ☐

3. EPI sequence (AP phase, ~3000 volumes ~ 25min), position to hit striatum and midbrain as shown ☐

4. After shim, before final gui: *We're now ready for your to play the game for 25mins it will be the same bleeping for the full duration.* ☐

5. *Just relax for 2 mins, we are just going to run another localiser, again there will be various noises"* ☐

6. Reverse phase (PA) fMRI (5 volumes) [If this hits stimulation limit > calculate> adjust rise time (rather than FOV)] ☐

8. AAhead_scout ☐

9. *"We're again ready for your to play the game for 25mins it will be the same bleeping for the full duration."* ☐

10. EPI sequence (AP phase encoding, 3000 volumes, 25min) ☐

11. Reverse phase (PA) fMRI (5 volumes) ☐

12. Get S out of scanner,15min break *"Do you need the toilet, or a drink?"* ☐

13. Read instructions & demo active phase ☐

**Data collection post-break**

0. Put S back in scanner – earplugs, phys noise, alarm bulb, key press box. ☐

1. Start game ☐

*2. "We now run a localiser scan to localise your head, It will only be 30s, expect buzzing, grinding, & knocking noises, nothing to do just relax"* ☐

3. AAhead_scout ☐

4. EPI sequence (AP phase, 3000 volumes) ☐

5. After shim & gui: *"We're now ready for your to play the game for 25mins, same bleeping noises"* ☐

6.*"Just relax for 2 mins, we are just going to run another localiser, again there will be various noises"* ☐

7. Reverse phase (PA) encoded fMRI (5 volumes) ☐

8. AAhead_scout ☐

9.*"We're now ready for your to play the game for a final 25mins, same bleeping noises"* ☐

10. EPI sequence (AP phase, 3000 volumes) ☐

11. Reverse phase (PA) fMRI (5 volumes)

*12. "Ok nearly there, we now have a final structural scan, where you have nothing to do except lie still for 6 minutes, and then we will get you out"* ☐

13. T1 – mprage1 ☐

**Get subject out of scanner**

Move cursor on test comp to empty file ☐

Debrief subject ☐

Inform accumulated wealth for the day, and if they are coming back it will be roughly the same again ☐

**Save data**

Behavioural data (mat files & txt files, both on USB + dropbox, place in data folder with txt files) ☐

Export whole folder (fMRI, T1, phys noise) to Samba ☐

Close session on both consoles ☐

Transfer to pacs > transfer > send to > DRCMR

Check complete > transfer > network job status

Check arrived at pacs > log in

**Clean up**

Tidy everything, make it spotless, (or wake with horse-head) ☐

Place bed half way down ☐

Put coil on shelf ☐

Put new paper on bed ☐

Put all equip in right place ☐

Wipe down surfaces with wetwipes ☐

Put phys monitors on charger, check if it charges ☐

**If final subject**

Return screen to shelf ☐

Turn off projector ☐

Check log books filled out ☐

Take all paper work with you ☐

Turn off all lights ☐

Place cushions into where headcoil went ☐

Close both computers & turn off ☐

Once both computers are turned off, press "System Off" under the quench button ☐

After pressing "System Off", turn the key to lock the scanner ☐

**Admin**

Put MR-safety protocol and subject code in locked drawer ☐

## Subject Instructions

**Introduction.** The experiment is divided over two days, within each day there will two different phases, a passive phase and an active phase. The main aim is to study how the brain reacts to changes in wealth. All of the money involved is real, and you will be will be paid out the total wealth accumulated, summed over the two days.

### Day 1

**Passive phase.** For the passive phase, you will see a number in the middle of the screen, this is your current wealth for the day in kr.

When you see a white box around the number, you are to press the button within 1s. (If you do not, you will be instructed to "press button earlier").

Shortly after pressing the button you will see an image in the background, and this will cause your wealth to change.

You are instructed to attend to any relationship between the images and the effect this has on your wealth, since in the active phase that follows you will be given the opportunity to choose images to influence your wealth.

<u>Learning these relationships can make a large difference to your earnings in the active phase.</u>

**Active phase**. With the money accumulated in the passive phase, you will play gambles composed of the same images.

In each trial, you will be presented with two of the images that you have learned about in the passive phase.

By pressing the buttons in the scanner to move a cursor, you now have the option to choose to either

   a) Accept gamble one, in which case you will be assigned one of the two images, each with 50% probability (not shown), or…

   b) Accept gamble two, in which case you will be assigned one of the two images, each with 50% probability (again not shown),.

The outcomes of your gambles will be hidden from you, and only 10 of them will be randomly chosen and applied to your current wealth.

You will be informed of your new wealth at the end of the active phase.

You can keep any money accumulated after the active phase.

If you do not choose in time, then we will give you one of the worst images, it is recommended that you always choose in time.

The decisions you make can make a big difference to your end wealth.

**Day 2**

**Introduction.** On day two, you will be endowed with a new wealth of 1000kr, and you will go through the same active and passive phases as described before, but the images will be new and they will be associated with different changes in wealth.

**At the end of the two days.** Your accumulated wealth will be added over the two days, and transferred to your account, within approximately two weeks, and is taxable under standard regulations (B-income).

Total earnings = (Wealth after day 1) + (Wealth after day 2)
This will be paid over and above your remuneration for participating in the experiment

**Data and code availability.** The datasets, analyses, stimuli, code, and codebook are available in the 'ergodicity-breaking-choice-experiment" repository: github.com/ollie-hulme/ergodicity-breaking-choice-experiment. All data figures have associated raw data. There are no restrictions on data availability.

**Acknowledgements.** We thank Ole Peters, Alex Adamou, Yonatan Berman, Mark Kirstein, Tobias Andersen, Jason Collins, Peter Dayan, Brad Cameron, Chris Merrill, Adam Goldstein, Alex Imas, Ilari Lehti, and Sven Resnjanskij for helpful discussions and comments. Thank you to Félix Hubert for making Supp. Fig. 6. Funding: HS (Lundbeck Foundation Grant of Excellence "ContAct" ref: R59 A5399 ; Novo Nordisk Foundation Interdisciplinary Synergy Programme Grant "BASICS" ref: NNF14OC0011413; 5-year professorship in precision medicine sponsored by Lundbeck Foundation ref: R186-2015-2138), O.J.H (Lundbeck Foundation, ref: R140-2013-13057; Danish Research Council ref: 12-126925), DM (Novo Nordisk Foundation project grant, ref: NNF16OC0023090).

**Contributions.** OH & DM conceived of the study, OH, FR, HS & DM designed the study, FR collected the data, OH & DM & MK performed modelling, OH performed analyses and wrote the paper. All authors contributed to the interpretation, edited, and approved the final version.

**Conflict of Interest.** HS has received honoraria as speaker from Sanofi Genzyme, Denmark and Novartis, Denmark, as consultant from Sanofi Genzyme, Denmark and as senior editor (NeuroImage) from Elsevier Publishers, Amsterdam, The Netherlands. HRS has received royalties as book editor from Springer Publishers, Stuttgart, Germany.